%% file: HofstedtPepper.tex
\documentclass{tlp}

\usepackage{epic,eepic}
\usepackage{boxedminipage}
\usepackage{amssymb,latexsym,stmaryrd}
\usepackage{theorem}
\usepackage{alltt}
\usepackage{xspace,url}
\usepackage{booktabs,ifthen}
\usepackage{pst-node}

\label{firstpage}

\title[Integration of Declarative and Constraint Programming]%
{Integration of Declarative and Constraint Programming}

\author[Petra Hofstedt and Peter Pepper]
         {PETRA HOFSTEDT and PETER PEPPER\\
         Technische Universit\"at Berlin, Sekr.~FR~5-13, 
         Franklinstr.~28/29, D-10587 Berlin\\
         \email{\{ph,pepper\}@cs.tu-berlin.de}}

\pagerange{\pageref{firstpage}--\pageref{lastpage}}

\usepackage{macros}

\submitted{18 December 2003}
\revised{20 February 2005, 25 November 2005}
\accepted{13 January 2006}

\begin{document}

\maketitle

\begin{abstract}
  Combining a set of existing constraint solvers into an integrated
  system of cooperating solvers is a useful and economic principle to
  solve hybrid constraint problems. In this paper we show that this
  approach can also be used to integrate different language paradigms
  into a unified framework. Furthermore, we study the syntactic,
  semantic and operational impacts of this idea for the amalgamation
  of declarative and constraint programming. \emph{To appear in Theory
    and Practice of Logic Programming (TPLP).}
\end{abstract}


\section{Introduction}

\input{files/Introduction}

\section{Declarative Programming Languages}
\label{sec:Declarative}

\input{files/Declarative}

\section{Cooperating Constraint Solvers}
\label{sec:CooperatingSolvers}

\input{files/Coop-solvers}

\section{Combination of Declarative and Constraint Programming}
\label{sec:Combination}

\input{files/Combination}

\section{Conclusion and Related Work}
\label{sec:Conclusion}

\input{files/Conclusion}


\bibliographystyle{acmtrans}
\bibliography{tplp-article}

\label{lastpage}

\end{document}

%% file: files/Introduction.tex
Declarative programming languages base on the idea that programs
should be as close as possible to the problem specification and
domain. In particular, the semantics of the computation does not
depend on the concepts of time and state.
Programs of these languages usually consist of directly formulated
mathematical objects, \ie of predicates and functions in logic and
functional (logic) languages \resp which are used to describe
properties of problems and required solutions.

Our point of view is to \emph{consider declarative programming as
  constraint programming:}
Syntactically this is evident. Logic languages are based on
predicates, and goals (on these predicates) are constraints.
For functional languages the underlying equality relations can be
regarded as constraints as well. But this point of view also applies
to the (operational) semantics: the evaluation of expressions in logic
and functional languages consists of their stepwise transformation to
a normal form, while particular knowledge is collected (in the form of
substitutions).  This corresponds to a stepwise propagation of
constraints.

This kind of consideration opens an interesting potential: In
\cite{Hofstedt_CP2000} a framework for cooperating constraint solvers
has been introduced which allows the integration of arbitrary solvers
and the handling of hybrid constraints.  Considering declarative
programming as constraint programming and looking at the language
evaluation mechanisms as constraint solvers we are able to integrate
these solvers into this framework.  Within the framework it is then
possible to extend the declarative languages by constraint systems
and, thus, to build constraint languages customised for a given set of
requirements for a comfortable modelling and solving of many problems.
The paper elaborates this approach.

We start with a recapitulation of necessary concepts and elements of
the syntax and semantics of declarative languages in
Sect.~\ref{sec:Declarative}.
Section~\ref{sec:CooperatingSolvers} reintroduces the frame\-work for
cooperating solvers of \citeN{Hofstedt_CP2000}.
In Sect.~\ref{sec:Combination} we examine the integration of a
functional logic language and consider the approach \wrt a logic
language.
We conclude our paper with a discussion of the gains and perspectives
of the approach in Sect.~\ref{sec:Conclusion} and compare it with
related work.




%% file: files/Declarative.tex

Declarative languages are roughly distinguished into functional,
logic, and constraint programming languages.

All declarative languages base on the concepts of signatures and terms.

\begin{definition}
    A \emph{signature} $\Sigma = (S,F,R)$ consists of a set $S$ of sorts, a set
    $F$ of $S$-sorted function symbols and a set $R$ of $S$-sorted predicate
    symbols.  $R$ contains in particular equality symbols for every sort
    \mbox{$s\in S$} and the predicate symbols $true$ and $false$.
    
    By ${\cal T}(F,X)$ we denote the usual set of \emph{free
      $F$-terms}, short: \emph{terms}, over the set $X$ of $S$-sorted
    variables. Variable-free terms are called \emph{ground terms}.
    Expressions $r(t_1,\ldots,t_n)$ with terms $t_i$ and a predicate
    symbol $r\in R$ as outermost symbol are called \emph{predicate terms}.
    By ${\cal T}(\Sigma,X)$ we denote the set of terms and predicate terms.

    Given a \emph{$\Sigma$-structure} $\cal D$ we obtain the usual notions of
    the value of a ground term $t$ in $\cal D$ and of the validity of a
    predicate term $p$ in $\cal D$, denoted as ${\cal D} \vDash p$. 
\end{definition}

In connection with the evaluation of programs the well-known notions of
substitutions and unifiers play a central role. We briefly recall their main
aspects.

\begin{definition}
    \label{def:Substitution}
    By $t[t^\prime]$ we denote a term $t$ with a distinguished subterm
    $t^\prime$. (This can be formally defined using either positions or
    contexts.)
    
    A \emph{substitution} $\sigma$ is a sort-preserving association
    $\{x_1=t_1,\ldots, x_m=t_m\}$ from variables $x_i\in X$ to terms
    $t_i\in {\cal T}(F,X)$.  (Since it fits more nicely into our overall
    framework and therefore simplifies some of the later
    presentations, we write substitutions here as special equations.)
    The application of a substitution $\sigma$ to a term or predicate
    term $e$ is denoted as $\sigma(e)$.
    
    A \emph{unifier} of two terms or predicate terms $t$ and
    $t^\prime$ is a substitution $\sigma$ which makes them equal:
    $\sigma(t) = \sigma(t^\prime)$. The most general unifier is
    denoted as $mgu(t,t^\prime)$.
    
    The \emph{composition} of some substitutions $\sigma$ and $\phi$
    is defined by $(\sigma\circ\phi)(x) = \sigma(\phi(x))$ for all
    $x\in X$. A substitution $\sigma$ is \emph{idempotent,} if
    $\sigma\circ\sigma = \sigma$ holds.
    
    The \emph{parallel composition $\uparrow$} of idempotent substitutions is
    defined as in \cite{Palamidessi}, \ie  $(\sigma \uparrow \phi) =
    mgu(\sigma, \phi)$. (Since we consider substitutions as special equations,
    $mgu$ is defined for them.)
\end{definition}

Starting from the foundation ${\cal T}(\Sigma,X)$ we build a hierarchy
of language paradigms.

\begin{center}
    \def\arraystretch{2.2}
    \begin{tabular}{ccc}
 \mc3{c}{\rnode{CFL}{constraint functional (logic)}} 
\\
 \rnode{FL}{functional logic} &  & {\rnode{CL}{constraint logic}}
\\
 \rnode{F}{functional} & \rnode{C}{constraint} & \rnode{L}{logic}
\\
 &  \rnode{T}{${\cal T}(\Sigma,X)$}
    \end{tabular}
    \psset{nodesep=3pt}
    \ncline{CFL}{FL}
    \ncline{CFL}{CL}
    \ncline{C}{CL}
    \ncline{L}{CL}
    \ncline{FL}{F}
    \ncline[linestyle=dashed]{FL}{L}
    \ncline{C}{T}
    \ncline{F}{T}
    \ncline{L}{T}
\end{center}

In the following we briefly sketch the syntax and the underlying ideas of this
hierarchy. In the subsequent sections we will then address the much more
important issues of the semantic and operational integration of the different
paradigms.

\input{files/Fp}

\input{files/Fll}

\input{files/Ll}

\input{files/Cp}

\input{files/Cl}

\input{files/Cfl}




%% file: files/Fp.tex

\paragraph{\bfseries Functional programming.} %
The realm of functional programming languages (examples are \Haskell
\cite{Haskell:99} and \Opal \cite{Opal:94}) is inhabited by a plethora
of syntactic variations.  For our conceptual treatment we can
constrain this to a minimal core that is as close as possible to the
other kinds of languages that we are treating here.
In the following, we distinguish two (disjoint) subsets of the set $F$
of function symbols of the signature $\Sigma$: the set
$\Delta\subseteq F$ of \emph{constructors} of the underlying
data types and the set $\Phi\subseteq F$ of \emph{defined functions}.

\begin{definition}
    A \emph{functional program} $P$ over $\Sigma$ is given by a finite set of
    rules of the form (called \emph{pattern-based definitions})

    \quad    $f(t_1,\ldots,t_n) \rightarrow t$

    \noindent%
    where $f\in \Phi$ is a defined function and the parameter terms
    $t_i\in{\cal T}(\Delta,X)$ are \emph{linear constructor terms},
    \ie are built up from constructors and variables such that every
    variable occurs only once. The right-hand side $t\in{\cal
      T}(F,X^\prime)$ is an arbitrary $F$-term, restricted to
    those variables $X^\prime\subseteq X$ that actually occur on the
    left-hand side.
\end{definition}

The \emph{evaluation} of a functional program $P$ reduces a given \emph{ground
  term} $e$ using the rules of $P$ until a normal form is obtained
\cite{FieldHarrison:88}. Each reduction step picks some function call
$f(e_1^\prime,\dots,e_n^\prime)$, that is, a subterm of
$e[f(e_1^\prime,\dots,e_n^\prime)]$, which can be unified with the left-hand
side of some rule $f(t_1,\ldots,t_n) \rightarrow t$. The resulting unifier $\sigma =
mgu(f(t_1,\ldots,t_n),f(e_1^\prime,\ldots,e_n^\prime))$ is then applied to the
right-hand side $t$ to derive the new term $e[\sigma(t)]$.

Since there may be different applicable rules for a chosen subterm caused by
overlapping left-hand sides, the rule selection strategy -- \eg first-fit or
best-fit -- of the language ensures a deterministic rule choice.

Moreover, there are different \emph{reduction strategies} for picking a redex in
each step, for example leftmost-innermost, leftmost-outermost, lazy, etc.
These strategies lead to quite different semantics as has already been
studied extensively in \cite{Manna:74}.
These differences are mainly reflected in the model-theoretic
interpretation of the equality $t_1 = t_2$ of the functional domain.
For the following illustrating examples an intuitive understanding of
equality will do. A thorough elaboration will be given in
Sect.~\ref{sec:Combination}.

\begin{example}%
\label{ex:addition}    
The following functional program provides rules for the addition of natural
numbers which are represented by the constructors {\tt 0} and {\tt s}.
\begin{tabbing}
{\tt add(s(X),Y)} \=$\rightarrow$ \={\tt s(add(X,Y))}\kill
{\tt add(0,X)} \>$\rightarrow$ \>{\tt X}             \hspace{60mm}\=(1)\\
{\tt add(s(X),Y)} \>$\rightarrow$ \>{\tt s(add(X,Y))}                   \>(2)
\end{tabbing}
\noindent%
This leads \eg to the following evaluation sequence, where the reduced
subterms are underlined:

\noindent%
{%
\def\lt#1{ $\leadsto_{(#1)}$ }%
\ttfamily%
\underline{add(s(0),s(s(0)))} \lt2 s(\underline{add(0,s(s(0)))}) \lt1 s(s(s(0)))
}
\end{example}

Since we will use it later on, we present a second example here. It comes from
the realm of resistors and their composition.

\begin{example}
For describing the composition of resistors we have three constructor functions:
a simple resistor, sequential composition, and parallel composition. The laws of
physics entail the program rules:
\begin{tabbing}
  $\rc(\Par(\R1,\R2))$ \=$\rightarrow$ \=$1 /(1 / \rc(\R1) + 1 / \rc(\R2))$\kill
  $\rc(\simple(\X))$   \>$\rightarrow$ \>$\X$ \\
  $\rc(\seq(\R1,\R2))$ \>$\rightarrow$ \>$\rc(\R1) + \rc(\R2)$ \\
  $\rc(\Par(\R1,\R2))$ \>$\rightarrow$ \>$1 /(1 / \rc(\R1) + 1 / \rc(\R2))$
\end{tabbing}
This allows \eg the following evaluation:

\smallskip
\noindent%
$\underline{\rc(\Par(\simple(300),\simple(600)))} \leadsto
 1 /(1 / \underline{\rc(\simple(300))} + 1 / \underline{\rc(\simple(600))})$
 
\vspace{1mm}
 
\noindent%
$ \leadsto 1 /(1 / 300 + 1 / 600) \leadsto 200$
\end{example}




%% file: files/Fll.tex
\paragraph{\bfseries Functional logic programming.}%
Functional logic programming was originally developed as an extension of
functional languages by concepts of logic languages, \cf
\cite{Reddy:85,Loogen:95}. However, they can as well be considered as embedding
functional concepts into a logic language, which is indicated by the dashed
connection in the above diagram. Typical representatives are \Babel \cite{Babel:1992}
and \Curry \cite{Curry:03}.

Syntactically, a functional logic program looks like a functional
program.  The difference lies in the evaluation mechanism. Whereas
functional programs only allow the reduction of ground terms to normal
forms, functional logic programs also allow the solving of equations
using residuation \cite{AitKaciNasr:89} and narrowing
\cite{Hanus_FLP94}.
    
A \emph{narrowing step} is a transition $e[l^\prime]
\leadsto_{l\rightarrow r, \sigma} \sigma(e[r])$, where $l \rightarrow r$ is a
rule from the program $P$ and $l^\prime$ is a non-variable term (\ie
$l^\prime\notin X$) such that $\sigma = \mathit{mgu}(l^\prime,l)$.

\begin{example}
  We again use the above addition example.  In order to solve the
  equation ${\tt add(s(A),B) = s(s(0))}$, we apply narrowing.  The
  chosen subterm is underlined.

\smallskip
\noindent%
{%
\def\lt#1#2{$\leadsto_{(#1),\{#2\}}$}%
\ttfamily%
\underline{add(s(A),B)} = s(s(0)) \lt{2}{\tt X1 = A, Y1 = B} s(\underline{add(A,B)}) = s(s(0)) 

\vspace{1mm}

\noindent%
\lt{1}{\tt A = 0, X2 = B} s(B) = s(s(0))

\smallskip
}

Thus, a solution of the initial equation is given by the substitution
$\sigma({\tt A}) = {\tt 0}$ and $\sigma({\tt B}) = {\tt s(0)}$
which is computed by unification of {\tt s(B)} and {\tt s(s(0))} after the
last narrowing step.
\end{example}

We need to ensure confluence of the rewrite system which is essential
for completeness. Thus, the rules of functional logic programs usually
must satisfy particular conditions, \eg linearity of the left-hand
sides, no free variables in the right-hand sides and (weak)
nonambiguity for lazy languages, \cf \cite{Hanus_FLP94}.

An overview of functional logic programming languages is given in
\cite{Hanus_FLP94}, where the following assessment is made:
``In comparison with pure functional languages, functional logic
languages have more expressive power due to the availability of
features like function inversion, partial data structures, and logic
variables.  In comparison with pure logic languages, functional logic
languages have a more efficient operational behaviour since functions
allow more deterministic evaluations than predicates.''

It should not come as a surprise that the aforementioned semantic
intricacies of functional languages carry over to functional logic
programming, leading to notions such as innermost narrowing and the
like, \cf \cite{Hanus_FLP94}. 
But there are more severe problems, which are often ignored in
the literature.  Consider the example from \cite{Hanus_FLP94}.
\begin{example}
  Consider the following program
  \begin{tabbing}
    $\tt f(X)$~\= $\rightarrow$~\= $\tt a$\\
    $\tt g(a)$~\> $\rightarrow$~\> $\tt a$
  \end{tabbing}
  \noindent%
  For the equation $\tt f(g(X))=a$ innermost narrowing yields the
  substitution $\{{\tt X = a}\}$
  as the only solution, whereas outermost narrowing provides the
  identity substitution $\{\}$ as solution. However, this depends on
  the language semantics. In most functional languages the function
  $\tt g$ would be considered undefined for all arguments but $\tt a$
  (since there are no patterns for the other cases). And in a
  call-by-value semantics this would entail that $\{{\tt X = a}\}$ is
  the only solution.  This illustrates that call-by-name or
  call-by-need semantics is incompatible with innermost strategies
  \cite{Manna:74}.  These observations will play a role in our later
  considerations in Sect.~\ref{sec:Combination}.
\end{example}



%% file: files/Ll.tex
\paragraph{\bfseries Logic programming.}

By contrast to functional programming, logic programming is based on
predicate terms.

\begin{definition}
    A \emph{logic program} $P$ is a set of rules of the form
    
    \quad $q_0(t_{0,1},\ldots,t_{0,m})\ {~\texttt{:-}~} \ q_{1}(t_{1,1},~\ldots,~t_{1,n}),
    \ldots, q_{k}(t_{k,1},\ldots,t_{k,r})$, \quad $k \geq 0$.

    \noindent%
    where the $q_i(\dots)$ are predicate terms and they are called
    \emph{atoms}.  The borderline case $k=0$, which has no conditions,
    is called a \emph{fact}.
\end{definition}

The \emph{evaluation} of a logic program is based on
\emph{resolution}. 
One starts with a \emph{goal} $(R_1 \wedge \dots \wedge R_l)$, which
is a conjunction of atoms, and adds its negation $(\neg R_{1} \vee
\ldots \vee \neg R_{l})$ to the set of rules.  Then one has to find a
\emph{refutation}, that is, a sequence of resolution steps which ends
with the empty clause {$\Box$}.

A \emph{resolution step} on a \emph{goal} $G = (R'_{1} \wedge 
\ldots \wedge R'_{m})$ and a program $P$ takes one subgoal $R'_i$ and a new
variant $r = \ (Q\ {\tt :-}\ Q_{1},\ldots,Q_{n}.)$, $n\geq0$, of a
rule of $P$ such that $R'_i$ and $Q$ can be unified with most general
unifier $\sigma$.  The result of the resolution step $G
\leadsto_{\sigma,r} G'$ is the new goal $G' = (\sigma(R'_{1}) \wedge
\ldots \wedge \sigma(R'_{i-1}) \wedge
\sigma(Q_{1}) \wedge \ldots \wedge \sigma(Q_{n}) \wedge
\sigma(R'_{i+1}) \wedge \ldots \wedge
\sigma(R'_{m}))$.

If a refutation can be computed, \ie $P \vDash \exists (R_{1} \wedge
\ldots \wedge R_{l})$ holds, then the computation yields an
\emph{answer substitution} $\phi$ (as composition of the unifiers
computed in the resolution steps) such that $P \vDash \forall \
\phi(R_{1} \wedge \ldots \wedge R_{l})$ holds.  For a detailed
description of logic programming and its well-known representative
\Prolog see for example \cite{Prolog}.



%% file: files/Cp.tex
\paragraph{\bfseries Constraint programming.}%

Here problems are specified by means of
\emph{constraints}, that is, first order formulas which
express conditions or restrictions describing properties of objects
and relations between them.  
Constraints come from constraint systems:

\begin{definition}  
  Let $\Sigma = (S, F, R)$ be a signature such that $R$ contains at
  least a predicate symbol \mbox{$=^{s}$} for every sort \mbox{$s\in
    S$.}  Let $X$ be a set of $\Sigma$-variables.
  Let ${\cal D}$ be a $\Sigma$-structure with equality, \ie for
  every predicate symbol \mbox{$=^{s}$} there is an according predicate
  which is an equivalence relation and fulfils the following requirements:

  \vspace{2mm}

  \noindent
  For all $f\in F$, $r\in R$ and all terms $t_{i},t'_{i}\in {\cal T}(F,X)$ of
  appropriate sorts $s_{i}$:

  \noindent
  If 
  for all $i$:
  ${\cal D} \vDash \forall (t_{i} =^{s_{i}} t'_{i})$, 
  then 
  \begin{itemize}  
  \item ${\cal D} \vDash \forall (f(t_{1},\ldots,t_{n}) =^{s}
    f(t'_{1},\ldots,t'_{n}))$, when $f(t_{1},\ldots,t_{n})$ and
    $f(t'_{1},\ldots,t'_{n})$ are both defined, or both terms are undefined,
    
  \item ${\cal D} \vDash \forall (r(t_{1},\ldots,t_{m})
    \leftrightarrow r(t'_{1},\ldots,t'_{m}))$, when $r(t_{1},\ldots,t_{n})$
    and $r(t'_{1},\ldots,t'_{n})$ are both defined, or both terms are undefined.
  \end{itemize}
  
A basic \emph{constraint} is of the form $r(t_{1},\ldots,t_{m})$, where
$r \in R$ and $t_{i}\in {\cal T}(F,X)$.
  The set of basic constraints over $\Sigma$ is denoted by ${\cal
    C}onstraint$.  It contains the two distinct
  constraints $true$ and $false$ with ${\cal D} \vDash true$ and
  ${\cal D} \nvDash false$.
  
  A \emph{constraint system} is a 4-tuple $\zeta = (\Sigma, {\cal D},
  X, {\cal C}ons)$, where $\{true,false\}\subseteq {\cal C}ons
  \subseteq {\cal C}onstraint$.
\end{definition}

\begin{example}
  A typical example are constraint systems for linear arithmetic
  with constraints that are equalities and inequalities, \eg
  $\X + 2*\Y = 7$ and $\X \leq 3.5$.
    
  In this realm the signature $\Sigma$ usually contains the function
  symbols $+, -, *, /$ and the relation symbols $=, \geq, \leq$.
\end{example}

The \emph{evaluation} of constraints is handled by \emph{constraint solvers}.
These are sophisticated algorithms for particular application domains, for
example the simplex algorithm for linear arithmetic.
Constraint solvers can not only check the satisfiability of
constraints but can also compute entailed constraints, projections and
even solutions.  A \emph{solution} of a constraint is a valuation
which satisfies it.

In order to allow more convenient programming, constraint systems can be
embedded into an appropriate language, which provides concepts like recursion,
encapsulation and abstraction. This leads to constraint logic and
constraint functional (logic) programming.




%% file: files/Cl.tex

\paragraph{\bfseries Constraint logic programming.}%
Logic programs are extended by constraints such that the right-hand
side of a rule may not only contain atoms but also constraints from an
arbitrary constraint system.
Consequently, the evaluation mechanism of logic languages, \viz resolution, has
to be extended by mechanisms for collecting constraints and checking their
satisfiability using appropriate constraint solvers
\cite{JaffarMaherMarriottStuckey}.

The first and also the most typical language which has been extended
by constraints, was the logic programming language \Prolog; the
initial motivation was to overcome the limitations of the expressive
power of the language when reasoning about arithmetic. 
A typical and comfortable constraint logic programming system is \Eclipse
\cite{Eclipse:03}.



%% file: files/Cfl.tex


\paragraph{\bfseries Constraint functional (logic) programming.} %
Functional (logic) languages can be extended further by
\emph{guarding} the rules with sets of constraints.

\begin{definition}\label{def:cfl}
    A \emph{constraint functional (logic) program} $P$ over $\Sigma$ is given by a
    finite set of rules of the form

    \quad    $f(t_1,\ldots,t_n) \rightarrow t \where G$

    \noindent%
    where -- as in functional logic programs -- $f\in \Phi$, $t_i\in{\cal
      T}(\Delta,X)$ and $t\in{\cal T}(F,X)$. But now we have in addition
      a set $G$ of (basic) constraints over $\Sigma$ and $X$.
\end{definition}

For example, \TOYFD \cite{FHgSp:PADL2003} and \TOYR
\cite{TOY-FD:1997} allow finite domain constraints and 
real arithmetic constraints, \resp.

An example of a constraint functional logic program and its evaluation
will be considered informally in Sect.~\ref{sec:Example}.
The evaluation mechanism is discussed in detail and formally in
Sect.~\ref{sec:Combination}.



%% file: files/Coop-solvers.tex
\begin{figure}
    \centering 
    \psset{unit=0.01\textwidth}
    \begin{tabular}[t]{c@{\hspace{1cm}}c}
     \input{files/Architecture1}
     &
     \input{files/Architecture2}
     \\
     (a) & (b)
    \end{tabular}
\caption{General architecture for (cooperating) constraint solvers}
\label{fig:cococos}
\end{figure}
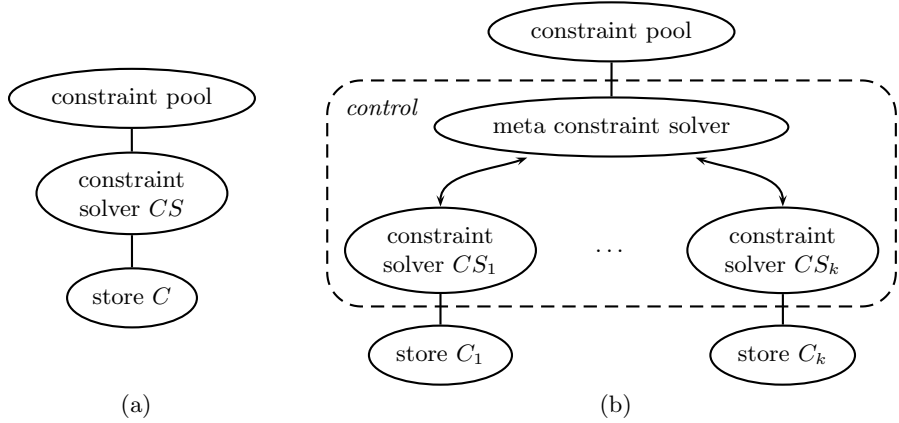

The basic architecture of a simple constraint solving system is a
\emph{solver algorithm} $CS$ associated to a \emph{constraint store}
$C$ and a \emph{constraint pool}; both are sets of (basic) constraints
(see Figure~\ref{fig:cococos}\,(a)).

\begin{itemize}
  \item Initially the constraint store of a solver is empty, more
    precisely: it contains only the constraint $true$; the constraint
    pool contains the constraints to solve.
  \item By the so-called \emph{constraint propagation} the solver adds
    constraints from the pool to its store while ensuring that the
    constraints in the store remain satisfiable.  In this way the set
    of possible valuations for the variables of the constraint store
    is successively narrowed.
  \item If the solver detects an inconsistency, the corresponding
    constraint is rejected.
\end{itemize}

When \emph{constraints from different realms} shall be used together,
one has two possibilities. Either one programs a new solver that is
capable of handling all kinds of constraints. Or one takes
several existing solvers, one for each realm, and coordinates them by
some mediating program. The former approach usually generates more
efficient solvers, but the amount of implementation work becomes
prohibitive, when more and more kinds of constraints shall be
integrated.  Therefore we focus on the second approach, which is more
flexible and more economic.

In \cite{Hofstedt_CL2000} a framework for cooperating constraint
solvers has been introduced and formally described, including
cooperation strategies for the solvers. In \cite{Dissertation}
termination and confluence, as well as soundness and completeness
restrictions are examined. An implementation of our system \METASl is
described in \cite{FHM03KI,FHM03FLAIRS}.

Figure~\ref{fig:cococos}\,(b) shows the architecture of our system for
cooperating solvers.  In the following let $L$ with $\mu,\nu\in L$
denote the set of indices of constraint systems.
\begin{itemize}
  \item The \emph{stores} $\C\nu$ of the individual constraint solvers
    $\CS{\nu}$ hold the constraints which have been propagated so far.
    Initially they are all empty.
  \item The \emph{constraint pool} is again the set of constraints
    that still need to be considered. Initially it contains the whole
    constraint problem to be solved.
  \item The \emph{meta solver} coordinates the work of the individual
    solvers.
        It distributes the constraints from the pool to the
        appropriate solvers, which put them into their stores by
        \emph{constraint propagation} and use them for their local
        computation (see the function $\tell{\nu}$ below).
%
        Conversely, constraints in the local stores may be
        \emph{projected} to the pool in order to make them available
        as new information to other solvers (see the functions
        $\proj{\nu}{\mu}$ below).
\end{itemize}
The process of sequent propagations and projections ends, when no more
information exchange takes place. Then the contents of the stores and
of the pool together represent the result:
it may indicate, whether the initial constraint conjunction was
unsatisfiable or not; moreover, restrictions of the solution space are
provided by means of projections of the stores. The restrictions may
even provide a full solution of the problem.

Details of the cooperation and communication of the involved solvers
are determined by the \emph{cooperation strategy} of the solvers.  A
cooperation strategy may influence the solution process with regard to
different criteria.
The solver cooperation system \METASl which implements our ideas
provides a flexible strategy definition framework.
One can prescribe particular search strategies, one can formulate
choice heuristics for constraints with respect to their gestalt and
their domains, one can specify the order of propagation and projection
and so forth. For this system it has been shown in \cite{FHM03KI} that
appropriate strategies for solver cooperation can yield comfortable
performance improvements for various kinds of constraint problems.

At the heart of our approach are the requirements for the
\emph{interfaces} by which the solvers are integrated into the system.
There are essentially two kinds of operations that constitute this
interface (\cf Sect.~\ref{sec:Propagation} and \ref{sec:Projection}):
\begin{itemize}
  \item For every solver $\CS{\nu}$ there is a function $\tell{\nu}$ 
    for propagating constraints from the pool to the store $\C{\nu}$.
  \item For every pair of solvers $\CS\nu$, $\CS\mu$ there is a
    function $\proj{\nu}{\mu}$ for providing information from the
    store $\C{\nu}$ to the solver $\CS{\mu}$ (via the constraint
    pool).  Note that this entails the translation into the other
    solver's signature.
\end{itemize}

\input{files/Example}

\input{files/Propagation}

\input{files/Projection}

\input{files/Semantics}




%% file: files/Architecture1.tex

\begin{pspicture}(0,0)(20,40)

\rput(10,30){\rnode{CP}{\psovalbox{\strut constraint pool\strut}}}

\rput(10,20){\rnode{CS}{\psovalbox{\parbox{5em}{\centering constraint solver $CS$}}}}

\rput(10,9){\rnode{C}{\psovalbox{\strut store $C$\strut}}}

\ncline{CS}{C}

\ncline{CP}{CS}

\end{pspicture}




%% file: files/Architecture2.tex

\begin{pspicture}(0,0)(60,40)

\rput(30,37){\rnode{Pool}{\psovalbox{\strut constraint pool\strut}}}

\rput(30,27){\rnode{Meta}{\psovalbox{\strut meta constraint solver\strut}}}

\rput(12,14){\rnode{CS1}{\psovalbox{\parbox{5em}{\centering constraint solver $CS_1$}}}}
\rput(30,14){\dots}
\rput(48,14){\rnode{CSk}{\psovalbox{\parbox{5em}{\centering constraint solver $CS_k$}}}}

\rput(12,03){\rnode{C1}{\psovalbox{\strut store $C_1$\strut}}}
\rput(48,03){\rnode{Ck}{\psovalbox{\strut store $C_k$\strut}}}

\psframe[fillstyle=none,framearc=.3,linestyle=dashed](00,32)(60,08)
\rput[lt](02,30){\emph{control}}

\ncline{Pool}{Meta}
\nccurve[angleA=200,angleB=90]{<->}{Meta}{CS1}
\nccurve[angleA=340,angleB=90]{<->}{Meta}{CSk}
\ncline{CS1}{C1}
\ncline{CSk}{Ck}

\end{pspicture}



%% file: files/Example.tex

\subsection{An Example}
\label{sec:Example}

To give a better intuition about the working of our approach, we present a small
example. It is taken from the realm of constraint functional logic programming
and illustrates the interaction of a functional logic language with a finite
domain solver and a solver for interval arithmetic.

\begin{example}\label{ex:gains.ll}
  The following program%
  \footnote{%
    We use so-called extra variables in the rules, \ie variables which
    occur in the body but not in the head. We discuss the issue of
    completeness in presence of extra variables briefly in
    Sect.~\ref{sec:FLLasCS}.  } describes resistors from a certain set
  $\{300\,\Omega$, \dots, $3000\,\Omega\}$, as well as the formulas
  for the sequential and parallel composition of resistors. The
  formulation is a mixture of functional logic programming and
  constraint programming. The first constraint uses the membership
  test $\in_{\cal FD}$ from a constraint system for finite domains,
  and the other two constraints use the equality $=_{\cal A}$ from a
  constraint system over rational arithmetic. The equality $=_{\cal
    FL}$ comes from the functional logic resolution mechanism.
  
\begin{tabbing}
$\rc(\Par(\R1,\R2))$ \=$\rightarrow \Z \where 1/\X+1/\Y =_{\cal A} 1/\Z, \quad\X=_{\cal FL}\rc(\R1), \quad\Y=_{\cal FL}\rc(\R2)$\kill
$\rc(\simple(\X))$ \>$\rightarrow \X \where \X\in_{\cal FD}\{300,600,900,1200, \ldots, 2700,3000\}$\\
$\rc(\seq(\R1,\R2))$ \>$\rightarrow \Z \where \X+\Y =_{\cal A} \Z, \quad\X=_{\cal FL}\rc(\R1), \quad\Y=_{\cal FL}\rc(\R2)$\\
$\rc(\Par(\R1,\R2))$ \>$\rightarrow \Z \where 1/\X+1/\Y =_{\cal A} 1/\Z, \quad\X=_{\cal FL}\rc(\R1), \quad\Y=_{\cal FL}\rc(\R2)$
\end{tabbing}

Note that the various subterms (including equations) in this program
are all homogeneous, that is, they are in ${\cal T}(\Sigma,X)$ for the
signature of one of the underlying solvers. This is possible in
general: by introducing auxiliary variables we can always turn hybrid
terms into homogeneous terms; this is called flattening of terms and
constraints, \resp  Again, this relies on an appropriate semantic
definition of the functional logic equality $=_{\cal FL}$ (which will
be discussed in Sect.~\ref{sec:Combination}).

In the following we sketch the way, in which our approach may handle
the above program. For the time being, the treatment is on a more
intuitive basis. The for\-ma\-li\-sation of the various steps will be given
subsequently. We use the following notation to illustrate the
snapshots from the evaluation, where the three stores $\C{\cal FL}$,
$\C{\cal FD}$ and $\C{\cal A}$ belong to the functional logic language
solver, the finite domain solver, and the arithmetic solver, \resp

\begin{Snapshot}
\Pool Constraint Pool
\Store1 \mathit{store~of}~\CS{\cal FL}
\Store2 \mathit{store~of}~\CS{\cal FD}
\Store3 \mathit{store~of}~\CS{\cal A}
\end{Snapshot}

\noindent%
(1) Suppose we want to compose two resistors in parallel and need a
combined resistance of $200\,\Omega$. The question is, which resistors
do we have to pick from our set. This question is formalizable in our
program as the equation \mbox{$\rc(\Par(\RA,\RB)) =_{\cal FL} 200$.}
This leads to the following initial configuration.

\begin{Snapshot}
\Pool \rc(\Par(\RA,\RB)) =_{\cal FL} 200
\Store1 \true
\Store2 \true
\Store3 \true
\end{Snapshot}

\noindent%
(2) We apply the third rule (using narrowing) and reach the following
system state.

\begin{Snapshot}
\Pool \Z =_{\cal FL} 200, \quad 1/\X + 1/\Y =_{\cal A} 1/\Z, \quad \X =_{\cal FL} \rc(\RA), \quad \Y  =_{\cal FL} \rc(\RB)
\Store1 \true
\Store2 \true
\Store3 \true
\end{Snapshot}

\noindent%
(3) When there are several constraints in the pool, the
particular cooperation strategy of the system decides, which
constraint to choose next for propagation, here \eg the goal $1/\X +
1/\Y =_{\cal A} 1/\Z$.  We propagate it (using the function
$\tell{\cal A}$) to the arithmetic solver $\CS{\cal A}$. This is
followed by a propagation of $\Z =_{\cal FL} 200$ to the store of $\CS{\cal
  FL}$. (In the pertinent stores we can omit the index of the equality
symbol.)

\begin{Snapshot}
\Pool \X =_{\cal FL} \rc(\RA), \quad \Y  =_{\cal FL} \rc(\RB)
\Store1 \Z = 200
\Store2 \true
\Store3 1/\X + 1/\Y = 1/\Z
\end{Snapshot}

\noindent%
(4) Next, the system chooses the goal $\X =_{\cal FL} \rc(\RA)$ for a
narrowing step based on the rules of our program (using the function
$\tell{\cal FL}$). This leads to a disjunction of three possibilities:

\begin{tabbing}
\quad  \= $\RA =_{\cal FL} \simple(\X) \wedge \X \in_{\cal FD}\{300,600,\ldots,3000\}$\\  
$\vee$ \> $\RA =_{\cal FL} \seq(\R1,\R2) \wedge (\X_1+\Y_1=_{\cal A} \X)
\wedge \X_1 =_{\cal FL} \rc(\R1) \wedge \Y_1 =_{\cal FL} \rc(\R2)$ \\
$\vee$ \> $\RA =_{\cal FL} \Par(\R1,\R2) \wedge (1/\X_2+1/\Y_2 =_{\cal A}
1/\X) \wedge \X_2 =_{\cal FL} \rc(\R1) \wedge \Y_2 =_{\cal FL} \rc(\R2)$
\end{tabbing}

\noindent 
Due to the disjunction, we have to form three instances of our configuration,
each representing one of the choices. For lack of space we only present the
derivation of the first alternative here. 

\begin{Snapshot}
\Pool \RA =_{\cal FL} \simple(\X), 
      \quad \X \in_{\cal FD}\{300,\ldots,3000\}, 
      \quad \Y  =_{\cal FL} \rc(\RB)
\Store1 \Z = 200
\Store2 \true
\Store3 1/\X + 1/\Y = 1/\Z
\end{Snapshot}
Note that the store $\C{\cal FL}$ did not change in this step. The
application of a program rule causes a replacement of the chosen
constraint by new ones according to the right-hand side of the rule.
In contrast, an enhancement of the store happens at the propagation of
constraints which do not contain defined functions any more, as we
will see in Step~(6).

\vspace{2mm}

\noindent%
(5) Now we apply the same process to the second goal $\Y =_{\cal FL}
\rc(\RB)$, again pursuing only the first variant.

\begin{Snapshot}
\Pool \RA =_{\cal FL} \simple(\X), 
      \X \in_{\cal FD}\{300,\ldots\},
      \RB =_{\cal FL} \simple(\Y), 
      \Y \in_{\cal FD}\{300,\ldots,3000\}  
\Store1 \Z = 200
\Store2 \true
\Store3 1/\X + 1/\Y = 1/\Z
\end{Snapshot}

\noindent%
(6) Next we propagate all the constraints of the pool to their associated
stores.

\begin{Snapshot}
\Pool 
\Store1 \vtop{
        \hbox{$\RA = \simple(\X)$}
        \hbox{$\RB = \simple(\Y)$}
        \hbox{$\Z = 200$\strut}
        }
\Store2 \vtop{
        \hbox{$\X \in\{300,\ldots,3000\}$}
        \hbox{$\Y \in\{300,\ldots,3000\}$}
        }
\Store3 \vtop{\hbox{$1/\X + 1/\Y = 1/\Z$}}
\end{Snapshot}

\noindent%
(7) At this point a system without solver cooperation would terminate the
computation and not draw any further conclusions.

But our system enables solver \emph{cooperation}. For the continuation of our
example we assume that the cooperation strategy of the system now
forces the finite domain solver to project its store by generating
bounds for the variables using the function $\proj{{\cal FD}}{{\cal A}}$.%
\footnote{%
  Actually, our implementation \METASl \cite{FHM03KI} distinguishes
  between weak projection generating only constraint conjunctions and
  strong projection which is allowed to project disjunctions as well
  (first proposed in \cite{Dissertation}).  Using different kinds of
  projections we are able to realise a variant of the Andorra
  principle \cite{CoWaYa:91,Warren:88} which proved to be very
  advantageous \wrt efficiency. The generation of bounds in our
  example represents a weak projection.\label{footnote:projections} }
This is followed by a projection of $\C{\cal FL}$ for the variable
$\Z$ common to $\C{\cal FL}$ and $\C{\cal A}$.

\begin{Snapshot}
\Pool   (\Z =_{\cal A} 200),\quad (300 \leq_{\cal A} \X),\quad (\X \leq_{\cal A} 3000),\quad (300\leq_{\cal A} \Y),\quad (\Y\leq_{\cal A} 3000)
\Store1 \vtop{
        \hbox{$\RA = \simple(\X)$}
        \hbox{$\RB = \simple(\Y)$}
        \hbox{$\Z = 200$\strut}
        }
\Store2 \vtop{
        \hbox{$\X \in \{300,\ldots,3000\}$}
        \hbox{$\Y \in \{300,\ldots,3000\}$}
        }
\Store3 \vtop{\hbox{$1/\X + 1/\Y = 1/\Z$}}
\end{Snapshot}

\noindent%
(8) The new constraints in the pool are now amenable to treatment by the arithmetic
solver. Therefore the meta solver propagates them (using $\tell{\cal A}$) to
this solver. Using its computational capabilities, the arithmetic solver can
derive more accurate bounds:

\begin{Snapshot}
\Pool   
\Store1 \vtop{
        \hbox{$\RA = \simple(\X)$}
        \hbox{$\RB = \simple(\Y)$}
        \hbox{$\Z = 200$\strut}
        }
\Store2 \vtop{
        \hbox{$\X \in \{300,\ldots,3000\}$}
        \hbox{$\Y \in \{300,\ldots,3000\}$}
        }
\Store3 \vtop{
        \hbox{$1/\X + 1/\Y = 1/200,\Z = 200$}
        \hbox{$(300\leq\X),~(\X\leq600)$}
        \hbox{$(300\leq\Y),~(\Y\leq600)$}
        }
\end{Snapshot}

\noindent%
(9) In the following steps the arithmetic solver's improved bounds can be projected
back to the pool, from where they are propagated to the finite domain solver,
which then narrows down its choices.

\begin{Snapshot}
\Pool   
\Store1 \vtop{
        \hbox{$\RA = \simple(\X)$}
        \hbox{$\RB = \simple(\Y)$}
        \hbox{$\Z = 200$\strut}
        }
\Store2 \vtop{
        \hbox{$\X \in \{300,600\}$}
        \hbox{$\Y \in \{300,600\}$}
        }
\Store3 \vtop{
        \hbox{$1/\X + 1/\Y = 1/200,\Z = 200$}
        \hbox{$(300\leq\X),~(\X\leq600)$}
        \hbox{$(300\leq\Y),~(\Y\leq600)$}
        }
\end{Snapshot}

\noindent%
(10) At this point we need strong projection (\cf footnote
(\ref{footnote:projections})) by which the finite domain solver puts a
disjunction of equations\\
\hbox to\textwidth{\hfill$(\X=_{\cal A} 300 \wedge \Y=_{\cal A} 300)~\vee~\ldots~\vee~(\X=_{\cal A} 600 \wedge \Y=_{\cal A} 600)$\hfill}\\
into the pool. Each of these four conjunctions can again be propagated
to the arithmetic solver. Two of them will lead to solutions, but the
other two propagations lead to inconsistencies in the arithmetic
solver and are therefore discarded.
One successful final configuration is:
\begin{Snapshot}
\Pool   
\Store1 \vtop{
        \hbox{$\RA = \simple(300)$}
        \hbox{$\RB = \simple(600)$}
        \hbox{$\Z = 200$}
        }
\Store2 \vtop{
        \hbox{$\X = 300$}
        \hbox{$\Y = 600$}
        }
\Store3 \vtop{
        \hbox{$\Z=200$}
        \hbox{$\X=300$}
        \hbox{$\Y=600$\strut}
        }
\end{Snapshot}


\noindent%
The solution $\{\RA = \simple(300), \RB = \simple(600)\}$ can be extracted from
the constraint store $\C{\cal FL}$ of the solver $\CS{\cal FL}$.
\end{example}

This small example already demonstrates the important role of
cooperation strategies for the efficiency of the computation.  For
example, the ability to control the order of weak and strong
projections allows a considerable restriction of the set of variable
assignments before an explicit search for solutions is initiated for
the remaining alternatives.
Suppose that we had applied the strong projection at an earlier point.
This would have caused a search across all 100 alternatives of
resistor combinations, \ie $(\X =_{\cal A} 300 \wedge \Y =_{\cal A}
300)$, \dots, $(\X =_{\cal A} 3000 \wedge \Y =_{\cal A} 3000)$.
As a matter of fact, it is in general a good strategy to delay the
introduction of disjunctions as long as possible for the sake of
efficiency. This is suggested by experiences with the KIDS system
\cite{WesSmi:01}.

Based on the intuitive insights provided by this example we will now
look more deeply into the precise definitions of the propagation
function $\tell{\nu}$ and the projection function $\proj{\nu}{\mu}$.




%% file: files/Propagation.tex

\subsection{Constraint Propagation ($\tell{\nu}$)}
\label{sec:Propagation}

As can be seen \eg in Step~(3) of the above example, the 
function $\tell{\nu}$, $\nu\in L$, takes a constraint $c \in {\cal C}ons_{\nu}$
(\ie a basic constraint of the constraint system of solver
$\CS{\nu}$) from the pool and adds it to the constraint store
$\C{\nu}$, which leads to a new store $\C{\nu}^\prime$.  There may
also be a remaining part $\cpp$ of $c$, which is put back into the
pool (but this happens rarely in practice).  

\begin{example}
  Suppose we have the constraint $\sqrt{\X} = \Y$ in the pool. This
  may be used to put the constraint $\X = \Y^2$ into the store of some
  solver, while keeping the constraint $\Y \geq 0$ in the pool.
\end{example}

\begin{figure}
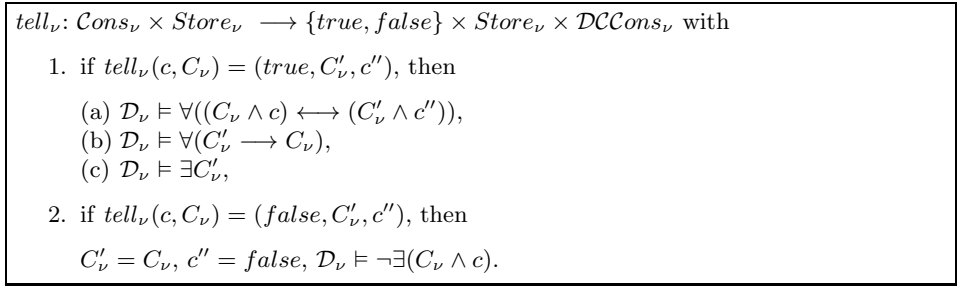

\noindent\begin{boxedminipage}{\linewidth}
$\tell{\nu}$: ${\cal C}ons_{\nu} \times Store_{\nu}\ \longrightarrow \{true,false\} \times Store_{\nu} \times {\cal DCC}ons_{\nu}$ with  
\begin{enumerate}
\item\label{tell.g1}
if $\tell{\nu}(c,\C{\nu}) = (true, \Cp{\nu}, \cpp)$, then 
\begin{tabbing}
 (a) \=${\cal D}_{\nu}\vDash \forall ((\C{\nu}\wedge c)\longleftrightarrow (\Cp{\nu}\wedge \cpp))$,\\ 
 (b) \>${\cal D}_{\nu} \vDash \forall (\Cp{\nu} \longrightarrow \C{\nu})$, \\
 (c) \>${\cal D}_{\nu} \vDash \exists \Cp{\nu}$,
\end{tabbing}
\item\label{tell.g2}
if $\tell{\nu}(c,\C{\nu}) = (false, \Cp{\nu}, \cpp)$, then 
\begin{tabbing}
$\Cp{\nu} = \C{\nu}$, $\cpp = false$, ${\cal D}_{\nu} \vDash \neg \exists (\C{\nu} \wedge c)$.
\end{tabbing}
\end{enumerate}
\end{boxedminipage}
\caption{Interface function $\tell{\nu}$, $\nu\in L$, (requirements)}\label{fig:tell_version3}
\end{figure}

Figure~\ref{fig:tell_version3} shows the requirements for the function
$\tell{\nu}$.%
\footnote{%
  In \cite{Hofstedt_CL2000} two forms of successful propagation are
  distinguished, which is necessary for general solvers to ensure
  termination of the system.  For the ``language solvers'' considered
  in this paper this is simplified.}
The function returns three values. The first one is a Boolean
indicator of the success or failure. The second one is the modified
store. And the third one is the remaining constraint $\cpp \in {\cal
  DCC}ons_{\nu}$, which is put back into the pool.
By ${\cal DCC}ons_{\nu}$ we denote the set of disjunctions of
constraint conjunctions.

When the solver successfully propagates a constraint $c$ to a store
$\C{\nu}$ \mbox{(Case~\ref{tell.g1}),} then
it must be ensured that the (overall) knowledge of the store and the
constraint is neither lost nor increased (a).
It is only possible to add constraints to a store, but not to delete
them.  Thus, the new constraint store $\Cp{\nu}$ must imply the old
one (b).
Of course, the new store $\Cp{\nu}$ has to be satisfiable in the
domain ${\cal D}_{\nu}$ of $\CS{\nu}$ as it is a constraint store (c).
In Example~\ref{ex:gains.ll}, \eg in Steps~(3) and (4), $\tell{\cal A}$
and $\tell{\FL}$ have been applied according to this definition.

This first case also covers the situation that a solver is not able to
handle a certain constraint $c$, \ie if the solver is incomplete. In
this case the store $\C{\nu}$ does not change and $c =c''$ remains in
the pool.%
\footnote{%
  Again, to ensure termination of the overall system, this particular
  case must be detected by the overall machinery and the treatment
  of the constraint must be suspended. We omit this technical detail
  in favour of readability.
}%

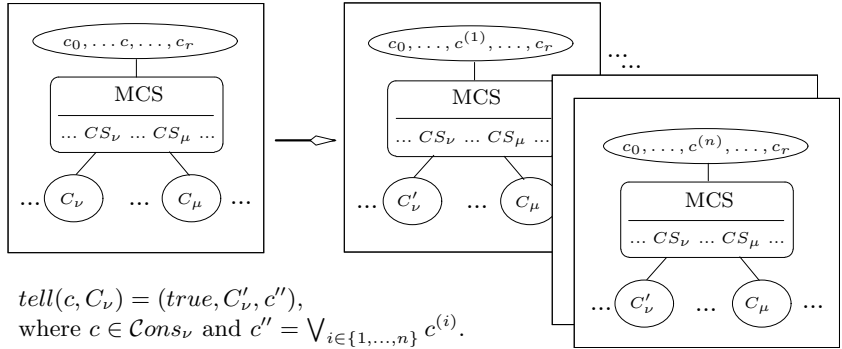
\begin{figure}
\centering
\let\SetFigFont=\undefined
\input{files/architecture.tell.v1.eepic} 
\let\SetFigFont=\undefined
\caption{Application of the interface function $tell$}\label{fig:tell}
\end{figure}

Figure~\ref{fig:tell} visualises the state change of the system when a
solver performs a successful constraint propagation.  The left side
shows the system before the propagation, the right side afterwards.
When we propagate $c$ to $\C{\nu}$ by $\tell{\nu}(c,\C{\nu})$, $c$ is
deleted from the pool and propagated to the store $\C{\nu}$.
The resulting new store $\Cp{\nu}$ and the remaining constraint $\cpp$ 
may in general be disjunctions of constraint conjunctions, \eg 
$\cpp = \bigvee_{i\in\{1,\ldots,n\}} c^{(i)}$.
Since store and pool are sets of basic constraints, this causes a
splitting of the system as shown in Figure~\ref{fig:tell}.

If $\tell{\nu}(c,\C{\nu})$ fails (Case~\ref{tell.g2}) because $c$ and $\C{\nu}$
are contradictory, then $false$ is added to the constraint
pool and $\C{\nu}$ does not change (not shown in Figure~\ref{fig:tell}).

\begin{example}
  The interface function $\tell{\cal A}$ of our rational arithmetic
  solver $\CS{\cal A}$ could work as follows (see \eg Step~(10) in
  Example~\ref{ex:gains.ll}):

Given the store $C = (1/\X+1/\Y =_{\cal A} 1/200)$
the propagation $\tell{\cal A}((\X =_{\cal A} 300), C) = (true,C^\prime,true)$ 
yields -- after some computation by the solver $\CS{\cal A}$ -- a new simplified
store $C^\prime = (\X =_{\cal A} 300 \wedge \Y =_{\cal A} 600)$.
On the other hand, 
$\tell{\cal A}((\Y >_{\cal A} 600), C^\prime) = (false,C^\prime,false)$ represents a
failing propagation.
\end{example}

For each concrete solver $\CS{\nu}$ the user must provide a suitable
function $\tell{\nu}$.
However, constraint propagation is mainly based on the satisfiability
test which is the main operation of a constraint solver. Moreover, the
requirements in Figure~\ref{fig:tell_version3} are chosen in such a
way that they allow an easy integration of many existing solvers into
the cooperating system, taking particular properties of solvers (like
their incompleteness (\cf Case~\ref{tell.g1}) or an existing
entailment test) into consideration.

Examples for such concrete propagation functions will be shown in
Sect.~\ref{sec:Combination} for the special cases of functional logic
and logic languages, when they are viewed as solvers.




%% file: files/architecture.tell.v1.eepic
\setlength{\unitlength}{0.00041667in}
\begingroup\makeatletter\ifx\SetFigFont\undefined%
\gdef\SetFigFont#1#2#3#4#5{%
  \reset@font\fontsize{#1}{#2pt}%
  \fontfamily{#3}\fontseries{#4}\fontshape{#5}%
  \selectfont}%
\fi\endgroup%
{\renewcommand{\dashlinestretch}{30}
\begin{picture}(10589,4422)(0,-10)
\whiten\path(7537,1225)(4287,1225)(4287,4395)
	(7537,4395)(7537,1225)
\path(7537,1225)(4287,1225)(4287,4395)
	(7537,4395)(7537,1225)
\whiten\path(3262,1218)(12,1218)(12,4388)
	(3262,4388)(3262,1218)
\path(3262,1218)(12,1218)(12,4388)
	(3262,4388)(3262,1218)
\put(5152,1870){\ellipse{750}{600}}
\put(6644,1873){\ellipse{750}{600}}
\put(4941,2548){\arc{210}{1.5708}{3.1416}}
\put(4941,3298){\arc{210}{3.1416}{4.7124}}
\put(6921,3298){\arc{210}{4.7124}{6.2832}}
\put(6921,2548){\arc{210}{0}{1.5708}}
\path(4836,2548)(4836,3298)
\path(4941,3403)(6921,3403)
\path(7026,3298)(7026,2548)
\path(6921,2443)(4941,2443)
\put(4931,2628){\makebox(0,0)[lb]{\smash{{\SetFigFont{6}{7.2}{\rmdefault}{\mddefault}{\updefault}... $CS_{\nu}$ ... $CS_{\mu}$ ...}}}}
\put(7519,1557){\makebox(0,0)[lb]{\smash{{\SetFigFont{10}{12.0}{\rmdefault}{\mddefault}{\updefault}...}}}}
\put(6459,1803){\makebox(0,0)[lb]{\smash{{\SetFigFont{6}{7.2}{\rmdefault}{\mddefault}{\updefault}$\C{\mu}$}}}}
\whiten\path(10297,305)(6927,305)(6927,3475)
	(10297,3475)(10297,305)
\path(10297,305)(6927,305)(6927,3475)
	(10297,3475)(10297,305)
\whiten\path(10577,12)(7207,12)(7207,3182)
	(10577,3182)(10577,12)
\path(10577,12)(7207,12)(7207,3182)
	(10577,3182)(10577,12)
\put(1629,3908){\ellipse{2600}{450}}
\put(2343,1907){\ellipse{750}{600}}
\put(9590,569){\ellipse{750}{600}}
\put(5892,3879){\ellipse{2600}{450}}
\put(8879,2571){\ellipse{2600}{450}}
\put(8098,568){\ellipse{750}{600}}
\put(848,1908){\ellipse{750}{600}}
\thicklines
\path(3415,2645)(4165,2645)
\whiten\thinlines
\path(3925.000,2585.000)(4165.000,2645.000)(3925.000,2705.000)(3853.000,2645.000)(3925.000,2585.000)
\path(1647,3675)(1647,3455)
\path(5939,3638)(5939,3418)
\path(8912,2354)(8912,2134)
\put(1362,3140){\makebox(0,0)[lb]{\smash{{\SetFigFont{8}{9.6}{\rmdefault}{\mddefault}{\updefault}MCS}}}}
\put(1542,1825){\makebox(0,0)[lb]{\smash{{\SetFigFont{10}{12.0}{\rmdefault}{\mddefault}{\updefault}...}}}}
\put(5654,3103){\makebox(0,0)[lb]{\smash{{\SetFigFont{8}{9.6}{\rmdefault}{\mddefault}{\updefault}MCS}}}}
\put(5834,1788){\makebox(0,0)[lb]{\smash{{\SetFigFont{10}{12.0}{\rmdefault}{\mddefault}{\updefault}...}}}}
\put(2835,1833){\makebox(0,0)[lb]{\smash{{\SetFigFont{10}{12.0}{\rmdefault}{\mddefault}{\updefault}...}}}}
\put(154,1817){\makebox(0,0)[lb]{\smash{{\SetFigFont{10}{12.0}{\rmdefault}{\mddefault}{\updefault}...}}}}
\put(626,3843){\makebox(0,0)[lb]{\smash{{\SetFigFont{6}{7.2}{\rmdefault}{\mddefault}{\updefault} $c_{0}, \ldots c, \ldots, c_{r}$}}}}
\put(4434,1786){\makebox(0,0)[lb]{\smash{{\SetFigFont{10}{12.0}{\rmdefault}{\mddefault}{\updefault}...}}}}
\put(7618,3702){\makebox(0,0)[lb]{\smash{{\SetFigFont{10}{12.0}{\rmdefault}{\mddefault}{\updefault}...}}}}
\put(7791,3567){\makebox(0,0)[lb]{\smash{{\SetFigFont{10}{12.0}{\rmdefault}{\mddefault}{\updefault}...}}}}
\put(8609,1802){\makebox(0,0)[lb]{\smash{{\SetFigFont{8}{9.6}{\rmdefault}{\mddefault}{\updefault}MCS}}}}
\put(8744,480){\makebox(0,0)[lb]{\smash{{\SetFigFont{10}{12.0}{\rmdefault}{\mddefault}{\updefault}...}}}}
\put(10105,510){\makebox(0,0)[lb]{\smash{{\SetFigFont{10}{12.0}{\rmdefault}{\mddefault}{\updefault}...}}}}
\put(7408,487){\makebox(0,0)[lb]{\smash{{\SetFigFont{10}{12.0}{\rmdefault}{\mddefault}{\updefault}...}}}}
\put(4840,3817){\makebox(0,0)[lb]{\smash{{\SetFigFont{6}{7.2}{\rmdefault}{\mddefault}{\updefault}$c_{0}, \ldots, c^{(1)}, \ldots, c_{r}$}}}}
\put(7816,2506){\makebox(0,0)[lb]{\smash{{\SetFigFont{6}{7.2}{\rmdefault}{\mddefault}{\updefault}$c_{0}, \ldots, c^{(n)}, \ldots, c_{r}$}}}}
\path(652,2965)(2622,2965)
\put(657,2600){\arc{210}{1.5708}{3.1416}}
\put(657,3350){\arc{210}{3.1416}{4.7124}}
\put(2637,3350){\arc{210}{4.7124}{6.2832}}
\put(2637,2600){\arc{210}{0}{1.5708}}
\path(552,2600)(552,3350)
\path(657,3455)(2637,3455)
\path(2742,3350)(2742,2600)
\path(2637,2495)(657,2495)
\path(1132,2495)(882,2215)
\path(4944,2928)(6914,2928)
\path(5424,2458)(5174,2178)
\path(2065,2501)(2315,2201)
\path(6318,2448)(6568,2148)
\path(7899,1627)(9869,1627)
\put(7904,1262){\arc{210}{1.5708}{3.1416}}
\put(7904,2012){\arc{210}{3.1416}{4.7124}}
\put(9884,2012){\arc{210}{4.7124}{6.2832}}
\put(9884,1262){\arc{210}{0}{1.5708}}
\path(7799,1262)(7799,2012)
\path(7904,2117)(9884,2117)
\path(9989,2012)(9989,1262)
\path(9884,1157)(7904,1157)
\path(9309,1167)(9559,867)
\path(8379,1157)(8129,877)
\put(147,535){\makebox(0,0)[lb]{\smash{{\SetFigFont{9}{10.8}{\rmdefault}{\mddefault}{\updefault}$tell(c,\C{\nu}) = (true, \Cp{\nu}, \cpp)$,}}}}
\put(147,160){\makebox(0,0)[lb]{\smash{{\SetFigFont{9}{10.8}{\rmdefault}{\mddefault}{\updefault}where $c\in {\cal C}ons_{\nu}$ and $\cpp = \bigvee_{i\in\{1,\ldots,n\}} c^{(i)}$.}}}}
\put(680,2671){\makebox(0,0)[lb]{\smash{{\SetFigFont{6}{7.2}{\rmdefault}{\mddefault}{\updefault}... $CS_{\nu}$ ... $CS_{\mu}$ ...}}}}
\put(7911,1332){\makebox(0,0)[lb]{\smash{{\SetFigFont{6}{7.2}{\rmdefault}{\mddefault}{\updefault}... $CS_{\nu}$ ... $CS_{\mu}$ ...}}}}
\put(4953,1815){\makebox(0,0)[lb]{\smash{{\SetFigFont{6}{7.2}{\rmdefault}{\mddefault}{\updefault}$\Cp{\nu}$}}}}
\put(7916,514){\makebox(0,0)[lb]{\smash{{\SetFigFont{6}{7.2}{\rmdefault}{\mddefault}{\updefault}$\Cp{\nu}$}}}}
\put(9370,510){\makebox(0,0)[lb]{\smash{{\SetFigFont{6}{7.2}{\rmdefault}{\mddefault}{\updefault}$\C{\mu}$}}}}
\put(674,1834){\makebox(0,0)[lb]{\smash{{\SetFigFont{6}{7.2}{\rmdefault}{\mddefault}{\updefault}$\C{\nu}$}}}}
\put(2167,1832){\makebox(0,0)[lb]{\smash{{\SetFigFont{6}{7.2}{\rmdefault}{\mddefault}{\updefault}$\C{\mu}$}}}}
\end{picture}
}

%% file: files/Projection.tex

\subsection{Projection of Constraint Stores ($\proj{\nu}{\mu}$)}
\label{sec:Projection}

Projection is used for information exchange between constraint solvers
$\CS{\nu}$ and $\CS{\mu}$, $\nu\not=\mu$, $\mu,\nu\in L$, via the
constraint pool.

Figure~\ref{fig:proj} shows the requirements for the function
$\proj{\nu}{\mu}$.
The function $\proj{\nu}{\mu}$ takes a set of common variables of the
two solvers and the local store $\C{\nu}$ and returns (a disjunction
of conjunctions of) constraints to the pool. 
$\proj{\nu}{\mu}(X, \C{\nu}) = c$ describes a projection of a store
$\C{\nu}$ \wrt common variables (\ie $X \subseteq X_{\nu} \cap
X_{\mu}$) to provide constraints $c$ of another solver $\CS{\mu}$.  It
provides knowledge implied by the store $\C{\nu}$.
Projection does not change the stores but only extends the pool by the
projected constraints.

To ensure that finally no solution is lost, a projection $c$ must
provide every satisfying valuation of the current store $\C{\nu}$.
That is, $\proj{\nu}{\mu}$ must be defined in such a way that every
solution $\sigma_{\nu}$ of $\C{\nu}$ is a solution of its projection
$c$ in ${\cal D}_{\mu}$ (soundness).

\begin{figure}
\noindent\begin{boxedminipage}{\linewidth}
$\proj{\nu}{\mu}:~\wp(X_{\nu} \cap X_{\mu}) \times Store_{\nu}
\rightarrow {\cal DCC}ons_{\mu}$, where ${\cal V}ars(\proj{\nu}{\mu} (X,\C{\nu})) = Y \subseteq X$, 
\begin{tabbing}
must be \emph{sound}, \ie for every valuation $\sigma_{\nu}$ for 
the variables of $Y$ must hold:\\[2mm]
If $({\cal D}_{\nu},\sigma_{\nu}) \vDash \exists\__{Y} \C{\nu}$, then 
    $({\cal D}_{\mu},\sigma_{\nu}) \vDash \proj{\nu}{\mu} (X,\C{\nu})$, where\\[1mm]

    $\exists\__{Y} \C{\nu}$ denotes the existential closure of
    formula $\C{\nu}$ except for the variables of $Y$.
\end{tabbing}
\end{boxedminipage}
\caption{Interface function $\proj{\nu}{\mu}$, $\mu\not=\nu$, $\mu,\nu\in L$, (requirements)}\label{fig:proj}
\end{figure}

\begin{example}\label{ex:example_5}
  Consider the arithmetic solver $\CS{\cal A}$ and the finite domain
  solver $\CS{\cal FD}$.
  Let $\C{\cal FD} = (\X \in_{\cal FD}\{300,\ldots,3000\} \wedge \Y
  \in_{\cal FD}\{300,\ldots,3000\})$ hold.  Define $\proj{\cal
    FD}{\cal A}$ (as a weak projection, \cf footnote
  (\ref{footnote:projections})) such that

\vspace{1mm}

\noindent
$\proj{\cal FD}{\cal A}(\{\X\},\C{\cal FD}) = ((\X\geq_{\cal A}
300)\wedge(\X\leq_{\cal A} 3000))$.

\vspace{1mm}

\noindent
This corresponds to Step~(7) of Example~\ref{ex:gains.ll}.
\end{example}

As in the case of $\tell{\nu}$ the function $\proj{\nu}{\mu}$ has to
be concretely programmed for every pair of given solvers. 
For many pairs of solvers it is possible to automatically provide
simple projection functions generating equality constraints which at
least express variable bindings. 
Often it is also sufficient to provide actual projection functions
only for particular pairs of solvers and to reduce superfluous
communication.

Examples of projection functions will also be shown in
Sect.~\ref{sec:Combination} in connection with constraint functional
logic and constraint logic languages.




%% file: files/Semantics.tex

\subsection{Semantics}
\label{sec:Semantics}

In \cite{Dissertation} we define reduction systems describing solver
collaborations based on the interface functions $\tell{\nu}$
and $\proj{\nu}{\mu}$.  The reduction systems work on so-called
configurations which consist of representations of the current
constraint pool and the associated stores, which together represent the state of
the system.

\citeN{Dissertation} gives a detailed discussion of the termination, confluence,
soundness and completeness of the reduction systems.

Based on the signatures and $\Sigma$-structures of the incorporated
constraint systems we build combined constraint systems by their
disjoint unions.  Clearly, for those sorts and function and predicate
symbols that are common to different systems the corresponding carrier
sets, functions and predicates must be identical. This applies in
particular to the semantics of the equality symbol (on shared sorts).

It is shown that -- using our method -- no solutions are lost.  If the system
fails then the given constraint problem is unsatisfiable.  Furthermore, we
discuss the situation of suspended constraints remaining in the pool due to the
incompleteness of solvers. If all incorporated solvers are complete and the
reduction relation is terminating, then for a satisfiable constraint problem the
method can be guaranteed to be successful. 

The soundness and completeness results do not depend on the particular
cooperation strategy of the solvers.

The (obvious) fact that the completeness of the overall system depends on the
completeness of the individual solvers will play a role in connection with
functional logic languages.



%% file: files/Combination.tex

As a particular application, our system of cooperating solvers allows the
integration of different host languages by treating them as constraint solvers.
As a matter of fact, it even makes it possible to work quite flexibly with
different evaluation strategies, that is, different operational semantics. A
good point in case are functional (logic) languages, which have very different
semantics depending on the chosen evaluation mechanisms. As mentioned in
Sect.~\ref{sec:Declarative}, this has already been
elaborated in the textbook \cite{Manna:74} for functional languages; for the
realm of functional logic programming the pertinent issues are described in
\cite{Hanus_FLP94,Hanus:95}.

The feasibility of this idea is based on two main observations: First, the
evaluation of expressions in declarative languages consists of their stepwise
transformation to a normal form while particular knowledge (in the form of
substitutions) is collected. Second, this way of proceeding is similar to a
stepwise propagation of constraints to a store, which is simplified in doing so.

In the following, we consider the integration of a functional logic
language and of a logic language, \resp, into the system of
cooperating solvers.
Syntactically, we extend the languages by constraints, but their evaluation
mechanisms are nearly unchanged: they are only extended by a mechanism for
collecting constraints of the other constraint solvers.

\paragraph{\bfseries A four-step process.}%
The integration of a declarative language into our system of cooperating
solvers requires four activities.
\begin{enumerate}
  \item The inherent constraints of the language have to be identified.
  \item Conversely, the constraints from the other domains have to be integrated
    into the syntax of the language.
  \item The language evaluation mechanism (\eg reduction or resolution) has to
    be extended by gathering constraints from the other domains.
  \item Finally one needs to carefully define the interface functions
    $\tell{.}$ and $\proj{.}{.}$ of the new language solver.
\end{enumerate}

\subsection{A Functional Logic Language as Constraint Solver}
\label{sec:FLLasCS}

\input{files/Fll_cs}

\subsection{A Logic Language as Constraint Solver}
\label{sec:LLasCS}

\input{files/Ll_cs}




%% file: files/Fll_cs.tex
The introduction of constraints into the rules of a functional logic
language yields constraint functional logic programming.
We follow our four step process. Recall that the basic syntactic
construct is

\quad    $f(t_1,\ldots,t_n) \rightarrow t \where G$

\noindent%
where $G$ is a set of constraints, including constraints from other
domains.  Moreover, all constraints in $G$ are homogeneous, that is,
are built from the signature of one solver, including equalities $t_1
=_{\FL} t_2$ of the functional logic language.

\paragraph{(1) Identifying language constraints.}

Functional logic languages are based on equalities between terms.  In
order to state precisely what this means, we have to look more closely
into the semantic models of these languages, \cf \eg
\cite{Manna:74,Winskel:93,BroyWirsingPepper:87}. It has to be ensured
that the \emph{operational semantics} induced by our cooperating
solvers is compatible with the \emph{mathematical semantics} of the
language under consideration.

Recall that the semantic value of a term $t$ in the model ${\cal D}$
is denoted by $\llbracket t\rrbracket_{\cal D}$; when ${\cal D}$ is
obvious from the context, we omit it and just write $\llbracket
t\rrbracket$.  
We are dealing with partial functions;
according to the traditional convention we write ``$\bot$''
(pronounced ``bottom'') to express partiality:
$\llbracket t\rrbracket = \bot$ means: $t$ has no proper
semantic value.
We note in passing that there are two kinds of partiality, both of
which can be subsumed under the element $\bot$. One is nontermination
of evaluations, the other is the lack of matching rules. For example,
if there is no rule for $\X/0$ -- independent of the responsible
solver -- then we have $\llbracket t/0\rrbracket = \bot$. The
following discussion is exemplified with nontermination but it applies
to both cases.


\begin{example}\label{ex:flatten}
  Consider the following superficial example over the natural numbers
  with constructors $0$ and $\succs$. (To ease reading we write
  $\succs$ without parentheses.)

\begin{tabbing}
$\f(\succs \X ,\Y)$ \=$\rightarrow \f(\X,\f(\succs \X,\Y))$\kill
$\f(0,\Y)$ \>$\rightarrow 0$\\
$\f(\succs \X ,\Y)$ \>$\rightarrow \f(\X,\f(\succs \X,\Y))$
\end{tabbing}
  We consider a functional logic solver $\CS{\FL}$ working in
  cooperation with others and we use again the notation from
  Example~\ref{ex:gains.ll} to illustrate the snapshots from the
  evaluation. The formalisation of the interface functions for
  $\CS{\FL}$ will be given and explained in more detail subsequently.

Start from the constraint 
$\Z =_{\cal FL} \f(0,\f(\succs 0,\succs 0))$
in the pool. 

{
\def\SnapshotSizeA{.40\textwidth}
\def\SnapshotSizeB{.25\textwidth}
\def\SnapshotSizeC{.25\textwidth}

\begin{Snapshot}
\Pool \Z =_{\cal FL} \f(0,\f(\succs 0,\succs 0))
\Store1 \true
\Store2 \true
\Store3 \true
\end{Snapshot}
}

In a \emph{call-by-name semantics} we first reduce the outermost $\f$
using the first rule of the program. This immediately yields the
result:
$\Z =_{\cal FL} 0$.
Semantically this means $\Sem{\Z} = \Sem{\f(0,\f(\succs 0,\succs
  0))}_{\mathit{cbn}} = \Sem{0}$.

In a \emph{call-by-value semantics} we first reduce the innermost $\f$
using the second rule:
$\Z =_{\cal FL} \f(0,\f(0,\f(\succs 0,\succs 0)))$.
As one can immediately see, this reduction process will never
terminate, that is, there will always be an equation $\Z =_{\cal FL}
t_i$ in the pool (with longer and longer right-hand sides $t_i$).
Semantically this means $\Sem{\Z} = \Sem{\f(0,\f(\succs 0,\succs
  0))}_{\mathit{cbv}} = \bot$.
\end{example}

What does this mean for the introduction of variables? Recall that we
need auxiliary variables for the homogeneity of the constraints for
different solvers, 
\ie we need to flatten constraints and terms. 
To see the pertinent problems, we modify the above example by
introducing several auxiliary variables. Moreover, we add a further
function $\g$.

\begin{tabbing}
$\f(\succs \X,\Y)$ \=$\rightarrow \f(\X,\W) \where \W =_{\cal FL} \f(\succs \X,\Y)$\kill
$\f(0,\Y)$ \>$\rightarrow 0$\\
$\f(\succs \X,\Y)$ \>$\rightarrow \f(\X,\W) \where \W =_{\cal FL} \f(\succs \X,\Y)$\\
$\g(\X)$ \>$\rightarrow 0 \where \f(\succs\X,\succs\X) =_{\cal FL} 0$
\end{tabbing}

Let us first consider the \emph{operational semantics}.  We start from

{
\def\SnapshotSizeA{.40\textwidth}
\def\SnapshotSizeB{.25\textwidth}
\def\SnapshotSizeC{.25\textwidth}

\begin{Snapshot}
\Pool \Z =_{\cal FL} \g(0)
\Store1 \true
\Store2 \true
\Store3 \true
\end{Snapshot}
After the first two steps this  leads to the following situation:
\begin{Snapshot}
\Pool \Z =_{\cal FL} 0,\ \f(0, \W) =_{\cal FL} 0,\ \W =_{\cal FL} \f(\succs 0,\ \succs 0)
\Store1 \true
\Store2 \true
\Store3 \true
\end{Snapshot}
In the next steps we may propagate $\Z =_{\cal FL} 0$ and evaluate the two calls
of $\f$:
\begin{Snapshot}
\Pool \ \W =_{\cal FL} \f(0, \W^\prime),\ \W^\prime =_{\cal FL} \f(\succs 0, \succs 0)
\Store1 \Z = 0
\Store2 \true
\Store3 \true
\end{Snapshot}
This may continue into the following configuration:
\begin{Snapshot}
\Pool \W^{(i)} =_{\cal FL} \f(\succs 0, \succs 0)
\Store1 \Z = 0,\ \W = 0,\ \W^\prime = 0,\ \dots
\Store2 \true
\Store3 \true
\end{Snapshot}
}
This process will obviously continue forever, creating more and more
auxiliary variables $\W^{(j)}$.

  Let us compare this \wrt the original (not flattened) definition of
  the function $\f$ in Example~\ref{ex:flatten}
and the two kinds of \emph{mathematical semantics} that we consider
here.
\begin{itemize}
  \item Under \emph{call-by-value} semantics, the result is $\Sem{\Z} =
    \Sem{\g(0)} = \bot$. 
    
  \item Under \emph{call-by-name} semantics the result is $\Sem{\Z} =
    \Sem{\g(0)} =\Sem{0}$.
\end{itemize}

Note that the auxiliary variables $\W$, $\W^\prime$, \dots do not
appear at all in the solution space of the mathematical semantics,
since they are only internal artefacts of the computation process.

Coming back to the operational semantics, we can only look up the
value of $\Z$ in the store, when the computation is finished, that is,
when the pool is empty. In the above example this will never happen;
therefore we will never be able to extract the value $\Z = 0$ from the
store. This means that $\Z$ has no value, \ie $\Sem{\Z} = \bot$. In
other words: Without further precautions, the introduction of
variables is only compatible with \emph{call-by-value} semantics.

How could we implement \emph{call-by-name} semantics and still allow
the introduction of variables? A simple solution consists of the
introduction of a dependency relation between variables. In our
example, $\W^\prime$ would depend on $\W$, because it occurs in the
term on the right-hand side of $\W$. As soon as the pertinent rule
would reduce $\W =_{\cal FL} \f(0, \W^\prime)$ to $\W =_{\cal FL} 0$,
the dependent variable $\W^\prime$ would be eliminated from the pool.
Moreover, the order of evaluation would have to follow the
dependencies.

As can be seen from this sketch, our method of handling hybrid
constraints by flattening is more naturally related to 
\emph{call-by-value} semantics, but with a little effort other
semantic principles such as \emph{call-by-name} can be accommodated as well.

We note in passing that these considerations do not only apply to
functional logic solvers, but to all kinds of solvers. Consider for
example an arithmetic solver with the rule $0 * \X = 0$. If we
multiply the term $\g(0)$ from the above example by $0$ we obtain from
$\Z = 0 * \g(0)$ the two constraints
$\Z = 0 * \X_1,\ \X_1 =_{\cal FL} \g(0)$, which leads to
$\Z = 0,\ \X_1 =_{\cal FL} \g(0)$.
Again, the differences between \emph{call-by-name} and \emph{call-by-value} can
be captured by introducing a dependency of the variable $\X_1$ on $\Z$.

\subparagraph{Remark on strict equality.}

From a semantic point of view it is unsatisfactory that we express
undefinedness only operationally by the fact that the pool does not
become empty. In a denotational setting, one would prefer to represent
this situation also as $\bot$. This brings strict equality into the
game.

In discussions about mathematical semantics there are usually two
kinds of equality: Under \emph{strong equality} we have \eg $(\bot
=_{\cal D} \bot) = true$ and $(\bot =_{\cal D} 3) = false$. And
\emph{strict equality} has the property $\llbracket t_1 =
t_2\rrbracket =_{\cal D} \bot$ if $\llbracket t_1\rrbracket =_{\cal D}
\bot$ or $\llbracket t_2\rrbracket =_{\cal D} \bot$.
Strong equality obeys the laws of classical two-valued logic,
but it is in general not decidable.
Strict equality is a ``normal'' operator in the language but the semantics leads
to all problems of three-valued logic.
Neither equality is ``better'' than the other, they just serve different
purposes.

\emph{Call-by-value} semantics evidently conforms nicely to strict
equality. In our above example we have $\Sem{\Z} = \Sem{\g(0)} =
\Sem{\f(\succs 0,\succs 0)} = \dots = \bot$. Therefore every pool
contains some equality of the kind
$\Sem{\W^{(i)} = \f(\succs 0, \succs 0)}$ and thus $\Sem{\W^{(i)} = \bot}$,
which -- under strict equality -- is $\bot$. Finally, we must
consider conjunctions as strict such that the whole configuration is
$\bot$. This exactly reflects the fact that our computation does not
terminate.%
\footnote{%
  The formal disjunctions that reflect the various branches of the evaluation
  need a different semantic treatment in order to accommodate the inherent
  nondeterminism.
} 
The overall system remains semantically consistent, since a ``wrong" equation
such as $\Z=0$ is in conjunction with $\bot$ and therefore does no harm. As a
matter of fact, the overall configuration just moves in each step from one
representation of the value $\bot$ to another representation of the value 
$\bot$.

The same considerations apply to \emph{call-by-name} semantics. Here
we have the situation that \eg $\Sem{\f(0,\bot)} = 0$. Therefore an
equality like $\f(0,\W) = 0$, where both terms are different from
$\bot$, has the same truth value under both kinds of equality.
However, if we encounter a function that is nonterminating even under
\emph{call-by-name} semantics, then we need again the strict equality
in order to represent the overall nontermination by $\bot$.

The fact that the operational semantics requires additional means such as
dependencies among variables is independent of the kind of equality. If one
considers the above example, the computation is consistent, since it simply
keeps adding valid equalities of the kind $\W^{(i)} = 0$ to the store. All we
need is a mechanism to stop it from doing this forever.

\paragraph{(2) Extending the language by constraints of other domains.}

The syntax of the language also contains the constraints from the
other domains (occurring in the part $\dots \where G$ of the rules).
Therefore they are also part of the language constraints.

To get a clean separation of concerns we also have to flatten hybrid
terms from different domains.
Suppose, for example, that we had given the third rule of our program
in the hybrid form

\begin{tabbing}
$\rc(\Par(\R1,\R2)) \rightarrow \Z \where 1/\rc(\R1)+1/\rc(\R2) =_{\cal A} 1/\Z$
\end{tabbing}

\noindent%
Then we would need to put this into the separated form

\begin{tabbing}
$\rc(\Par(\R1,\R2)) \rightarrow \Z \where 1/\X+1/\Y =_{\cal A} 1/\Z, \quad\X=_{\FL}\rc(\R1), \quad\Y=_{\FL}\rc(\R2)$
\end{tabbing}

\paragraph{(3) Extending the language evaluation mechanism by gathering
  constraints.}

Functional logic languages are based on \emph{narrowing}. Therefore we
consider this evaluation mechanism as constraint solver $\CS{\FL}$.
But in addition to performing the narrowing steps, the solver must
also collect constraints from other domains (occurring in the part
$\dots \where G$).

The basic concept of a \emph{narrowing step with constraints} can be
described as follows: Let $e$ be an equation with a distinguished
non-variable subterm $t$, that is, $e[t]$ with $t\notin X$. Let $(l
\rightarrow r \where G)$ be a rule from the program such that $\sigma
= mgu(t,l)$ unifies the subterm $t$ with the left-hand side $l$ of the
rule. Then the narrowing step yields the new equation $\sigma(e[r])$
together with the rewritten constraint $\sigma(G)$,
we write: $e[t] \leadsto_{\sigma} (\sigma(e[r]), \sigma(G))$.

The extensive discussion of the previous pages has shown the various
possibilities for choosing evaluation strategies and their impacts on
the semantics.

\paragraph{(4) Defining the interface functions $\tell{\FL}$ and
  $\proj{\FL}{\nu}$ of the particular language solver.}

This final step has to integrate the effects of narrowing and
constraint collection with the other solvers.

\begin{figure}
\centering
\input{files/architecture.fll.eepic}
\caption{Application of the interface function $\tell{\FL}$ (see also 
  Figure~\ref{fig:tell_FL})}
\label{fig:tell_FL_picture}
\end{figure}
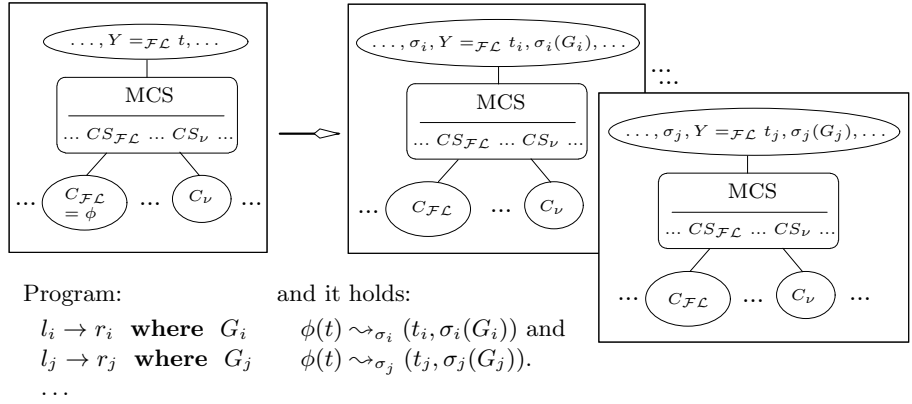

\paragraph{Propagation.}%
Figure~\ref{fig:tell_FL_picture} illustrates how the interface
function $\tell{\FL}$ is used to simulate a narrowing step with
constraints, and Figure~\ref{fig:tell_FL} gives the formal definition
of the pertinent requirements.
To ease the presentation we suppose that all functional logic
equalities are in the form $Y =_{\FL} t$ with variable $Y$ and term
$t$. This can always be achieved with the help of auxiliary variables.

Like all solvers, $\CS{\FL}$ propagates constraints to its
store $\C{\FL}$ thereby checking the satisfiability of $\C{\FL}$ in
conjunction with the new constraints.  Therefore the function
$\tell{\FL}$ incorporates the principle of narrowing. 

In the light of the preceding discussion on flattening we assume that
all constraints which get into the pool (either as part of the initial
problem to solve or as results of propagations or projections) are
decomposed with the help of auxiliary variables such that all subterms
of the form $f(t_1, \dots, t_n)$, where $f$ is a defined function, are
extracted.
The definition of $tell_{\FL}$ therefore only needs to consider
narrowing steps on outermost terms. 
%
The distinction between \emph{call-by-value} and \emph{call-by-name}
must therefore be based on the aforementioned dependency relation
among variables.

We have to distinguish two kinds of constraints $Y =_{\FL} t$ (see
Figure~\ref{fig:tell_FL}):

\mbox{(\ref{fl.1})} When the term $t$ still contains defined
functions, a narrowing step is applied as part of $\tell{\FL}$. This
is only reflected by a change of the constraint pool. The store does
not change in this case. Note that due to the 
flattening $t$ contains exactly one defined function $f$, and this
function is the outermost symbol in $t$.  
Moreover, since the substitution $\phi$ defined in
Figure~\ref{fig:tell_FL} only contains constructor terms,
$\that=\phi(t)$ retains this property.

\mbox{(\ref{fl.2})} When the term $t$ is a constructor term then the
constraint $Y =_{\FL} t$ is added to the store if possible. Thereby,
the satisfiability test is realised by the parallel composition
$\uparrow$ of substitutions.
 
This integration of the narrowing into the propagation sometimes leads
to earlier bindings of variables and thus to a faster recognition of
unsuccessful computations. The price to be paid is that the solver
$\CS{\FL}$ is amalgamated into the function $\tell{\FL}$ -- and with
it all its problems.

%
\begin{figure}
\noindent\begin{boxedminipage}{\linewidth}

$\tell{\FL}$:
Let $P$ be a functional logic program with constraints, let \mbox{$\C{\FL} =
  \phi$} be the current constraint store of $\CS{\FL}$.  (Recall that the
constraint store $\C{\FL}$ is nothing but a substitution $\phi$ from variables
to constructor terms, which is written in the form of equations and thus can be treated like
constraints.)  Let $c = (Y =_{\FL} t)$ be the constraint to be propagated.  Let
$\that=\phi(t)$.

\vspace{1mm}

Finally, we use the following notion: A rule $p= (l_p \rightarrow r_p
\where G_p)$ \emph{applies to} $\that$, if for $\that\not\in X$ there
is a unifier $\sigma_p = mgu(\that,l_p)$.

\vspace{1mm}

We need to distinguish the following cases

\begin{enumerate}
  \item\label{fl.1} Let $\that$ contain defined functions; that is, $\that$ is
    of the form $f(\dots)$ with $f$ being the only defined function
    (due to the maximal flattening).  If the set $P_{c} \subseteq P$ of
    applicable rules is nonempty, then

    $\tell{\FL}(c, \C{\FL}) =$
    $(true, ~\C{\FL},~ \bigvee_{p \in P_{c}}
    (\sigma_p \wedge (Y =_{\FL} \sigma_p(r_p)) \wedge \sigma_p(G_p)))$.

\vspace{1mm}

\item\label{fl.2}
  Let $\that$ be a constructor term, \ie $\that \in {\cal
    T}(\Delta,X_{\FL})$.

\begin{enumerate}
\item\label{fl.2a}
  If $(\{Y = t\} \uparrow \C{\FL}) \not= \emptyset$, then
  
  $\tell{\FL}(c, \C{\FL}) = (true,~\{Y = t\} \uparrow \C{\FL}~,
  true)$.

\vspace{1mm}

\item\label{fl.2b}  
  If $(\{Y = t\} \uparrow \C{\FL}) = \emptyset$, then

  $\tell{\FL}( c, \C{\FL}) = (false , \C{\FL}, false)$.
\end{enumerate}

\end{enumerate}

\end{boxedminipage}
\caption{Interface function $\tell{\FL}$}\label{fig:tell_FL}
\end{figure}


\begin{example}
  To elucidate the interface definition we use our running example
  from Sect.~\ref{sec:Example}. In Step~(2) of this example we applied
  the only matching rule

\begin{tabbing}
$\rc(\Par(\R1,\R2)) \rightarrow \Z \where 1/\X+1/\Y =_{\cal A} 1/\Z, \quad\X=_{\FL}\rc(\R1), \quad\Y=_{\FL}\rc(\R2)$
\end{tabbing}

\noindent%
to the initial configuration. We split the equation in the pool into
two equations using an auxiliary variable such that the pool reads:

\Store{$\Z_1 =_{\cal FL} \rc(\Par(\RA,\RB)), \quad \Z_1 =_{\cal FL} 200$}

\noindent%
We pick the first term and apply $\tell{\FL}$ as described in
Figure~\ref{fig:tell_FL}. Note that we have the special situation, where the
store $\C{\FL}=\phi$ is still empty such that $\that = \phi(\rc(\Par(\RA,\RB)))
= \rc(\Par(\RA,\RB))$.

The narrowing step for the term with (a new instance of) the rule
unifies the subterm $\rc(\Par(\RA,\RB))$ -- which happens to be the
full term -- with the left-hand side $\rc(\Par(\R1,\R2))$ of the rule.
Since this matching exists, we have Case~\ref{fl.1} of
Figure~\ref{fig:tell_FL}.  The resulting most general unifier is the
substitution $\sigma = \{\R1=\RA,\,\R2=\RB\}$. Applied to the rule
this substitution yields the instantiated right-hand side $\Z$ and
constraints $1/\X+1/\Y =_{\cal A} 1/\Z, ~\X=_{\FL}\rc(\RA),
\Y=_{\FL}\rc(\RB)$. This is put back into the pool:

\Store{$\R1=_{\cal FL}\RA,~\R2=_{\cal FL}\RB,~\Z_1=_{\cal FL}\Z,~1/\X + 1/\Y
  =_{\cal A} 1/\Z,~\X =_{\cal FL} \rc(\RA),$ $\Y =_{\cal FL} \rc(\RB),~\Z_1 =_{\cal FL}200$}

If there is more than one applicable rule, we get a number of
newly built constraint pools.

Applying $\tell{\FL}$ to the two constraints $\R1=_{\cal FL}\RA$ and $\R2=_{\cal FL}\RB$
leads to Case~\ref{fl.2a} of Figure~\ref{fig:tell_FL}. As a result the
store $\C{\FL}$ then contains these constraints (a substitution).  The
remaining constraints in the pool are as follows:

\Store{$\Z_1=_{\cal FL}\Z,~\Z_1 =_{\cal FL} 200,~1/\X + 1/\Y =_{\cal A} 1/\Z,~\X =_{\cal FL} \rc(\RA),~\Y =_{\cal FL} \rc(\RB)$.}
Note, that Example~\ref{ex:gains.ll} displays simplified forms for brevity.
\end{example}
 
\paragraph{Projection.}%
Since the constraint store $\C{\FL}$ only contains substitutions, the
projection is trivial. It generates equality constraints representing
them.
For example, in Step~(7) the constraint $\Z =_{\cal A} 200$ is transferred from
$\C{\FL}$ to the pool using $\proj{\cal FL}{\cal A}$.

The specification of the function $\proj{\cal FL}{\nu}$ is given in
Figure~\ref{fig:proj_FL}.
(Note, that the equalities in the result $\phi|_X$ are indexed ${\nu}$
to express the fact that they now belong to the domain of the solver
$\CS{\nu}$.)

\begin{figure}
\noindent\begin{boxedminipage}{\linewidth}

$\proj{{\cal FL}}{\nu}$:
The projection of a store $\C{\cal FL} = \phi$ \wrt a constraint
system $\zeta_\nu$ and a set of variables $X \subseteq X_{\cal
  FL}\cap X_{\nu}$ makes the substitutions for ${\tt x} \in X$ explicit:

\vspace{1mm}

\noindent
$\proj{{\cal FL}}{\nu}(X, \phi) = 
 \cases{
  \phi|_X &
     \begin{minipage}[t]{0.4\linewidth}
       if $\phi\not=\emptyset$
     \end{minipage} 
  \cr     
  true & 
     otherwise.}$
\end{boxedminipage}
\caption{Interface function $\proj{{\cal FL}}{\nu}, \nu\in L$}
\label{fig:proj_FL}
\end{figure}

\paragraph{Completeness and other aspects.}%
Since the interface functions $\tell{\cal FL}$ and $\proj{\cal FL}{\nu}$ of our
functional logic language solver fulfil the requirements given in
Sect.~\ref{sec:CooperatingSolvers}, the soundness and completeness results of
the cooperation framework hold for the integration of a functional logic
language -- however, only relative to the completeness of the narrowing strategy
encoded within $\tell{\FL}$.

It is well known that narrowing is in principle a complete and sound
method for theories presented by confluent and terminating rewrite
systems. But in connection with functional logic languages and in
particular in the presence of extra variables the issue of
completeness becomes more intricate. This problem has been
investigated by several authors, among others
\cite{MidHam:94,SMI:95,Hanus:95}. These papers discuss slightly
different syntactic criteria, which guarantee that certain narrowing
strategies (\eg lazy or needed narrowing) are sound and complete.

Evidently, these criteria carry over to the narrowing strategy that we
embed into our function $\tell{\FL}$. For lack of space we cannot go
into details here. Suffice it to say that, for example, our program
rule

\begin{tabbing}
$\rc(\Par(\R1,\R2)) \rightarrow \Z \where 1/\X+1/\Y =_{\cal A} 1/\Z, \quad\X=_{\FL}\rc(\R1), \quad\Y=_{\FL}\rc(\R2)$
\end{tabbing}

\noindent%
contains the three extra variables $\X$, $\Y$ and $\Z$ and is in the
class 3-CTRS of \cite{MidHam:94}. But the syntactic form also meets
the requirements of being \emph{orthogonal}, \emph{properly oriented}
and \emph{right-stable} \cite{SMI:95}, which guarantees that the
program is level-confluent. (These conditions essentially state that
the variables are in a nice left-to-right order.)
    
As a matter of fact, the rule is also \emph{constructor-based} and
\emph{functional} in the sense of \cite{Hanus:95}, since the variables
$\R1$ and $\R2$ in the guarding constraints occur in the left-hand
side and the variable $\Z$ of the right-hand side is defined by the
guarding constraints. This also entails the completeness of lazy and
needed narrowing.  
But Hanus imposes a further requirement: the
equality $=_{\FL}$ has to be strict (which allows to consider it as an
equality ${(Y \equiv t )} =_{\FL} true$).  Since we only put
substitutions with constructor terms into the store $\C{\FL}$, this
works in our approach as well. (This is similar to the language
\Curry \cite{Curry:03}.)

However, there is one further issue! The above quoted requirements
refer to the functional logic equalities in the guard $G$. But our
rules have a more general form:

\begin{tabbing}
  $f(t_1,\dots, t_n) \rightarrow r \where c_1, \dots, c_m, u_1=_{\FL}v_1, \dots, u_k
  =_{\FL} v_k$
\end{tabbing}
    
\noindent%
Only the equalities $u_i=_{\FL}v_i$ (with $u_i$ being a constructor
term, possibly only a variable) belong to the functional logic solver.
The $c_j$ are constraints from other domains.
    
This raises the question: How do these constraints $c_1, \dots, c_m$
from the other domains fit into the picture? Let us consider our
standard example

\begin{tabbing}
$\rc(\Par(\R1,\R2)) \rightarrow \Z \where 1/\X+1/\Y =_{\cal A} 1/\Z, \quad\X=_{\FL}\rc(\R1), \quad\Y=_{\FL}\rc(\R2)$
\end{tabbing}
     
\noindent%
Without the first constraint, the variable $\Z$ would not occur in the
guard and thus the rule would not even be 3-CTRS (and thus no
completeness results could be given at all). On the other hand, all
constraints $c_i$ do not contain any of the defined functions and
therefore do not participate in any kind of narrowing strategy. So,
from the point of view of narrowing, a variable like $\Z$ can be
treated like a constant.
     
If one of the other solvers finds a solution, \eg $\Z = 200$ in our
example, then the corresponding substitution is put into the store
$\C{\FL}$. This way the associated value is guaranteed to be a ground
constructor term and thus does not interfere with the narrowing
strategies.

This illustrates again the -- evident -- fact that the completeness of
the overall system, and thus also of its narrowing part, depends on
the completeness of all participating solvers.




%% file: files/architecture.fll.eepic
\setlength{\unitlength}{0.00041667in}
\begingroup\makeatletter\ifx\SetFigFont\undefined%
\gdef\SetFigFont#1#2#3#4#5{%
  \reset@font\fontsize{#1}{#2pt}%
  \fontfamily{#3}\fontseries{#4}\fontshape{#5}%
  \selectfont}%
\fi\endgroup%
{\renewcommand{\dashlinestretch}{30}
\begin{picture}(11454,5111)(0,-10)
\whiten\path(3292,1914)(12,1914)(12,5084)
	(3292,5084)(3292,1914)
\path(3292,1914)(12,1914)(12,5084)
	(3292,5084)(3292,1914)
\whiten\path(8092,1899)(4322,1899)(4322,5069)
	(8092,5069)(8092,1899)
\path(8092,1899)(4322,1899)(4322,5069)
	(8092,5069)(8092,1899)
\whiten\path(11442,769)(7505,769)(7505,3939)
	(11442,3939)(11442,769)
\path(11442,769)(7505,769)(7505,3939)
	(11442,3939)(11442,769)
\put(6937,2506){\ellipse{750}{600}}
\put(10136,1368){\ellipse{750}{600}}
\put(9951,1298){\makebox(0,0)[lb]{\smash{{\SetFigFont{6}{7.2}{\rmdefault}{\mddefault}{\updefault}$\C{\nu}$}}}}
\put(6752,2436){\makebox(0,0)[lb]{\smash{{\SetFigFont{6}{7.2}{\rmdefault}{\mddefault}{\updefault}$\C{\nu}$}}}}
\put(1749,4589){\ellipse{2600}{450}}
\put(971,2542){\ellipse{1080}{700}}
\put(5337,2461){\ellipse{1080}{700}}
\put(6222,4573){\ellipse{3600}{600}}
\put(9480,3444){\ellipse{3700}{600}}
\put(8639,1319){\ellipse{1080}{700}}
\put(2463,2588){\ellipse{750}{600}}
\path(1767,4356)(1767,4136)
\thicklines
\path(3452,3421)(4202,3421)
\whiten\thinlines
\path(3962.000,3361.000)(4202.000,3421.000)(3962.000,3481.000)(3890.000,3421.000)(3962.000,3361.000)
\path(6232,4271)(6232,4051)
\path(9395,3138)(9395,2918)
\put(1482,3821){\makebox(0,0)[lb]{\smash{{\SetFigFont{8}{9.6}{\rmdefault}{\mddefault}{\updefault}MCS}}}}
\put(1662,2506){\makebox(0,0)[lb]{\smash{{\SetFigFont{10}{12.0}{\rmdefault}{\mddefault}{\updefault}...}}}}
\put(2954,2513){\makebox(0,0)[lb]{\smash{{\SetFigFont{10}{12.0}{\rmdefault}{\mddefault}{\updefault}...}}}}
\put(87,2513){\makebox(0,0)[lb]{\smash{{\SetFigFont{10}{12.0}{\rmdefault}{\mddefault}{\updefault}...}}}}
\put(5947,3736){\makebox(0,0)[lb]{\smash{{\SetFigFont{8}{9.6}{\rmdefault}{\mddefault}{\updefault}MCS}}}}
\put(6127,2421){\makebox(0,0)[lb]{\smash{{\SetFigFont{10}{12.0}{\rmdefault}{\mddefault}{\updefault}...}}}}
\put(4441,2406){\makebox(0,0)[lb]{\smash{{\SetFigFont{10}{12.0}{\rmdefault}{\mddefault}{\updefault}...}}}}
\put(8156,4200){\makebox(0,0)[lb]{\smash{{\SetFigFont{10}{12.0}{\rmdefault}{\mddefault}{\updefault}...}}}}
\put(8256,4046){\makebox(0,0)[lb]{\smash{{\SetFigFont{10}{12.0}{\rmdefault}{\mddefault}{\updefault}...}}}}
\put(10687,1297){\makebox(0,0)[lb]{\smash{{\SetFigFont{10}{12.0}{\rmdefault}{\mddefault}{\updefault}...}}}}
\put(9146,2598){\makebox(0,0)[lb]{\smash{{\SetFigFont{8}{9.6}{\rmdefault}{\mddefault}{\updefault}MCS}}}}
\put(9326,1283){\makebox(0,0)[lb]{\smash{{\SetFigFont{10}{12.0}{\rmdefault}{\mddefault}{\updefault}...}}}}
\put(7744,1298){\makebox(0,0)[lb]{\smash{{\SetFigFont{10}{12.0}{\rmdefault}{\mddefault}{\updefault}...}}}}
\put(702,4546){\makebox(0,0)[lb]{\smash{{\SetFigFont{6}{7.2}{\rmdefault}{\mddefault}{\updefault} $\ldots, Y =_{\cal FL} t, \ldots$}}}}
\put(4617,4516){\makebox(0,0)[lb]{\smash{{\SetFigFont{6}{7.2}{\rmdefault}{\mddefault}{\updefault}$\ldots, \sigma_i, Y =_{\cal FL} t_i, \sigma_i(G_i), \ldots$}}}}
\put(7842,3391){\makebox(0,0)[lb]{\smash{{\SetFigFont{6}{7.2}{\rmdefault}{\mddefault}{\updefault}$\ldots, \sigma_j, Y =_{\cal FL} t_j, \sigma_j(G_j), \ldots$}}}}
\path(772,3646)(2742,3646)
\path(2182,3186)(2432,2886)
\path(1252,3176)(1002,2896)
\put(727,3281){\arc{210}{1.5708}{3.1416}}
\put(727,4031){\arc{210}{3.1416}{4.7124}}
\put(2807,4031){\arc{210}{4.7124}{6.2832}}
\put(2807,3281){\arc{210}{0}{1.5708}}
\path(622,3281)(622,4031)
\path(727,4136)(2807,4136)
\path(2912,4031)(2912,3281)
\path(2807,3176)(727,3176)
\path(5237,3561)(7207,3561)
\path(6647,3101)(6897,2801)
\path(5717,3091)(5467,2811)
\put(5170,3201){\arc{210}{1.5708}{3.1416}}
\put(5170,3951){\arc{210}{3.1416}{4.7124}}
\put(7250,3951){\arc{210}{4.7124}{6.2832}}
\put(7250,3201){\arc{210}{0}{1.5708}}
\path(5065,3201)(5065,3951)
\path(5170,4056)(7250,4056)
\path(7355,3951)(7355,3201)
\path(7250,3096)(5170,3096)
\path(8436,2423)(10406,2423)
\path(9846,1963)(10096,1663)
\path(8916,1953)(8666,1673)
\put(8369,2063){\arc{210}{1.5708}{3.1416}}
\put(8369,2813){\arc{210}{3.1416}{4.7124}}
\put(10449,2813){\arc{210}{4.7124}{6.2832}}
\put(10449,2063){\arc{210}{0}{1.5708}}
\path(8264,2063)(8264,2813)
\path(8369,2918)(10449,2918)
\path(10554,2813)(10554,2063)
\path(10449,1958)(8369,1958)
\put(192,1291){\makebox(0,0)[lb]{\smash{{\SetFigFont{9}{10.8}{\rmdefault}{\mddefault}{\updefault}Program:}}}}
\put(417,91){\makebox(0,0)[lb]{\smash{{\SetFigFont{9}{10.8}{\rmdefault}{\mddefault}{\updefault}$\ldots$}}}}
\put(417,466){\makebox(0,0)[lb]{\smash{{\SetFigFont{9}{10.8}{\rmdefault}{\mddefault}{\updefault}$l_j \rightarrow r_j \ \where \ G_j$}}}}
\put(3717,841){\makebox(0,0)[lb]{\smash{{\SetFigFont{9}{10.8}{\rmdefault}{\mddefault}{\updefault}$\phi(t) \leadsto_{\sigma_i} (t_i,\sigma_i(G_i))$ and}}}}
\put(3717,466){\makebox(0,0)[lb]{\smash{{\SetFigFont{9}{10.8}{\rmdefault}{\mddefault}{\updefault}$\phi(t) \leadsto_{\sigma_j} (t_j,\sigma_j(G_j))$.}}}}
\put(3417,1291){\makebox(0,0)[lb]{\smash{{\SetFigFont{9}{10.8}{\rmdefault}{\mddefault}{\updefault}and it holds:}}}}
\put(417,841){\makebox(0,0)[lb]{\smash{{\SetFigFont{9}{10.8}{\rmdefault}{\mddefault}{\updefault}$l_i \rightarrow r_i \ \where \ G_i$}}}}
\put(695,3351){\makebox(0,0)[lb]{\smash{{\SetFigFont{6}{7.2}{\rmdefault}{\mddefault}{\updefault}... $CS_{\cal FL}$ ... $CS_{\nu}$ ...}}}}
\put(5143,3280){\makebox(0,0)[lb]{\smash{{\SetFigFont{6}{7.2}{\rmdefault}{\mddefault}{\updefault}... $CS_{\cal FL}$ ... $CS_{\nu}$ ...}}}}
\put(5127,2425){\makebox(0,0)[lb]{\smash{{\SetFigFont{6}{7.2}{\rmdefault}{\mddefault}{\updefault}$\C{\cal FL}$}}}}
\put(8342,2142){\makebox(0,0)[lb]{\smash{{\SetFigFont{6}{7.2}{\rmdefault}{\mddefault}{\updefault}... $CS_{\cal FL}$ ... $CS_{\nu}$ ...}}}}
\put(719,2583){\makebox(0,0)[lb]{\smash{{\SetFigFont{6}{7.2}{\rmdefault}{\mddefault}{\updefault}$\C{\cal FL}$}}}}
\put(717,2341){\makebox(0,0)[lb]{\smash{{\SetFigFont{6}{7.2}{\rmdefault}{\mddefault}{\updefault}$= \phi$}}}}
\put(2292,2566){\makebox(0,0)[lb]{\smash{{\SetFigFont{6}{7.2}{\rmdefault}{\mddefault}{\updefault}$\C{\nu}$}}}}
\put(8386,1257){\makebox(0,0)[lb]{\smash{{\SetFigFont{6}{7.2}{\rmdefault}{\mddefault}{\updefault}$\C{\cal FL}$}}}}
\end{picture}
}

%% file: files/Ll_cs.tex
The integration of a logic language into the system of cooperating solvers
yields a constraint logic programming language. Since this is actually simpler than
the case of functional logic languages, we only sketch it here briefly.

\paragraph{(1 \& 2) Identifying language constraints and extending 
  the language by constraints of other domains.}

It is widely accepted that logic programming can be interpreted as
constraint programming over the Herbrand universe.  The appropriate
constraint solving mechanism $\CS{\cal L}$ is resolution.

The goals according to a given constraint logic program $P$ are the
natural constraints of a logic language solver. Furthermore, the set
${\cal C}ons_{\cal L}$ of constraints of this solver must contain
equality constraints $Y =_{\cal L} t$ between variables and terms to
represent substitutions.%
\footnote{Not all CLP systems support equalities on the syntactical
  level.  Rather they only generate them internally in the solver.  In
  our system, they are explicitly visible.}

We extend the syntax of the language by constraints of other
constraint systems which yields the typical CLP syntax,
\cf \eg \cite{JaffarLassezCLP}.  Thus, the set ${\cal C}ons_{\cal
  L}$ must furthermore include all constraints of the incorporated
solver(s).

\paragraph{(3 \& 4) Extending the language evaluation mechanism by 
  gathering constraints and defining the interface functions of the
  particular language solver.}

For the integration of $\CS{\cal L}$ into the system the interface
functions $\tell{\cal L}$ and $\proj{{\cal L}}{\nu}$ must be defined.
Step (3), \ie gathering constraints during resolution, is realised by
the extension of the resolution step from atoms to the whole body
including the constraints of other domains.

\paragraph{Propagation.}
The propagation function $\tell{\cal L}$ emulates resolution steps
(including gathering constraints). Its formal definition is given in
Figure~\ref{fig:tell_logic}.
Case~\ref{1a} represents a resolution step on a goal $R$ as a
successful propagation, where for every applicable rule we get a newly
created constraint pool and, thus, a new instantiation of the
architecture.
If there is no applicable rule for a goal, \ie Case \ref{1b}, the
propagation fails (in contrast to the functional logic solver
considered before, undefinedness of a predicate is regarded as
failure here).

Similar to the definition of $tell_{\cal FL}$ the remaining cases
describe the propagation of equality constraints by parallel
composition of substitutions (Case~\ref{2}).

\paragraph{Projection.} 
As before, the projection function provides constraints representing
the substitution from the store which has been computed during
resolution.
The definition of the projection function $\proj{\cal L}{\nu}$ is the
same as for the functional logic language solver given in
Figure~\ref{fig:proj_FL}, where the index ${\cal FL}$ is replaced by
${\cal L}$ to denote the origin of the projection.


\begin{figure}
\begin{boxedminipage}{\linewidth}
$\tell{\cal L}$:
Let $P$ be a constraint logic program, let \mbox{$\C{\cal L} = \phi$}
be the current store of $\CS{\cal L}$.

\begin{enumerate}
  \item\label{1} Let $R = p(t_{1},\ldots,t_{m})$ be the constraint
    (goal) which is to be propagated.  Let $\hat{R} = \phi(R)$.
    We use the following notion: A rule $p= (Q_p ~\texttt{:-}\ rhs_p)$
    \emph{applies to} $\hat{R}$, if there is a unifier $\sigma_p =
    mgu(\hat{R},Q_p)$.

    \begin{enumerate}
      \item\label{1a} If the set $P_R \subseteq P$ of applicable
        rules is nonempty, then
    
        $\tell{\cal L}(R, \C{\cal L}) = (true,~\C{\cal
          L},~\bigvee_{p\in P_{R}} (\sigma_p\wedge
        \sigma_{p}(rhs_p)))$.

\vspace{1mm}
    
      \item\label{1b}
        If there is no applicable rule in $P$, then
        
        $\tell{\cal L}(R, \C{\cal L}) = (false, \C{\cal L}, false)$.
    \end{enumerate}

\vspace{1mm}
    
  \item\label{2} Let $c = (Y =_{\cal L} t)$ be the constraint which
    is to be propagated.

    \begin{enumerate}
      \item\label{2a}
        If $(\{Y = t\} \uparrow \C{\cal L}) \not= \emptyset$, then

        $\tell{\cal L}(c, \C{\cal L}) = (true,~\{Y = t\} \uparrow \C{\cal L}~, true)$.

\vspace{1mm}
    
      \item\label{2b}
        If $(\{Y = t\} \uparrow \C{\cal L}) = \emptyset$, then 

        $\tell{\cal L}(c, \C{\cal L}) = (false , \C{\cal L}, false)$.
    \end{enumerate}
  \end{enumerate}
\end{boxedminipage}

\caption{Interface function $\tell{\cal L}$}\label{fig:tell_logic}
\end{figure}

Since the interface functions $\tell{\cal L}$ and $\proj{\cal L}{\nu}$
fulfil the requirements given in Sect.~\ref{sec:CooperatingSolvers},
the soundness and completeness results of the cooperation framework
hold for the integration of a logic language.




%% file: files/Conclusion.tex

This paper describes a general approach for the integration of
declarative languages and constraint systems. This essentially means
to treat their evaluation mechanisms together with programs as
constraint solvers.  This is done in \emph{four general steps}:

\begin{enumerate}
  \item Identifying language constraints,
  \item Extending the language by constraints of other domains,
  \item Extending the language evaluation mechanism by gathering
    constraints,
  \item Defining the interface functions of the particular language
    solver.
\end{enumerate}

The most important aspect of our approach is that the overall system
for cooperating solvers allows the handling of hybrid constraints over
different domains.


\paragraph{Gains and perspectives.}


Our general framework for cooperating solvers provides mechanisms for
the definition of cooperation strategies.
Similar to CLP(X) \cite{JaffarLassezCLP} and CFLP(X)
\cite{Lopez:ALP1992}, which are covered by our approach, the framework
can thus be instantiated by three parameters: a strategy definition
${\cal S}$, a set ${\cal X}$ of constraint systems and a host language
${\cal Y}$.
In this way, our approach enables the \emph{building of constraint languages
  customised for a given set of requirements for a comfortable modelling and the
  solution of many problems.}

For example, if one needs a convenient language to express search problems
with special constraints over finite and/or further domains, the user
chooses a logic language and according constraint systems. For
problems which allow or require a more deterministic modelling one may
decide to build a constraint language based on a functional or
functional logic language. In a similar way one can imagine to combine
a database language with constraints \cite{ConstraintDB:1995} or a
particular special-purpose language or system, such as an expert system,
a geographical information system, or a planning system. Of course,
the user is responsible for a sound specification of the language
constraints and the definition of the interface functions.


The meta-solver system \METASl \cite{FHM03FLAIRS,FHM03KI} implements
our ideas. Even though \METASl provides programming constructs for its
integration into other applications, which may even be imperative
languages, the true integration of imperative languages according to
our approach is an open question and a topic of future research.  The
main reason is that declarative languages abstract from real-world
issues such as time and state while imperative languages are
time-dependent which complicates their integration also for our
approach.

\vspace{2mm}


The choice of an appropriate cooperation strategy plays an important
role for the efficiency of the cooperating system \cite{FHM03KI}.
 
The definition of solver cooperation strategies is also very
interesting in the case of language solvers. This will
allow for example to switch from depth-first search
to breadth-first search or an evaluation mechanism based on the Andorra
principle \cite{CoWaYa:91,Warren:88} using a logic language without
reimplementing the evaluation mechanisms. The system \METASl already
offers predefined strategy patterns for these search strategies which
can be refined by problem-dependent knowledge or user knowledge about
the program.

If the user is able to define cooperation strategies for the solvers
and the language(s) or to refine them on the base of predefined
strategy patterns (as given in our implementation), she/he can also
employ problem-dependent knowledge and user knowledge, \eg about the
termination of particular predicates, to guide the computation.

First results \cite{FHR:04} on the integration of language solvers
into the meta-solver framework \METASl according to the described
approach confirm our theoretical considerations.

\vspace{2mm}


Finally, the approach opens a further interesting perspective:
A simple approach for the combination of different languages consists
of the definition of an explicit interface and providing language
constructs for initialising subcomputations in the particular
languages.  In this way one reaches a loose coupling and
interaction of programs written in different languages.

Our cooperation framework bases on a similar idea: It provides a meta
mechanism which takes care of the strategy of the cooperating solvers
and provides the constraint pool for maintaining and managing common
data and constraints. Besides this the main concept is the uniform
solver interface which allows to integrate declarative languages as
solvers as shown above. Using this interface it is not only possible
to integrate constraint solvers and language evaluation mechanisms but
also to integrate language evaluation mechanisms among each other
by appropriate interface definitions. This finally yields a language
interaction according to the above sketched interface model.


\paragraph{Related Work.}

\citeN{Lopez:ALP1992} considered a general scheme CFLP(X) for
constraint functional logic programming. The scheme is based on lazy
functional logic languages and allows beside conditions constraints in
the guards of the rules.
Using our cooperation approach, we achieve a covering of the CFLP(X)
and CLP(X) \cite{JaffarLassezCLP} approaches.  
In \cite{TOY-FD:1997} and \cite{FHgSp:PADL2003}, extending CFLP(X),
functional logic programming is integrated with real arithmetic and
finite domain constraints respectively.
The lazy functional logic languages \TOYR and \TOYFD are the
respective implementations.
\citeN{Lux:FLOPS2001} integrates linear constraints over real
numbers into the functional logic language \Curry in a similar way. 

\OpenCFLP by \citeN{KobayashiMarinIda:2003} combines a functional
logic host language with collaborating equational solvers which may be
distributed in an open environment. It provides the user with a
declarative strategy definition for the cooperating solvers basing
upon a set of basic operators.
However, the strategy language of \METASl gives finer control over the
individual collaboration steps because of its well considered solver
interface on the one hand and its structural pattern-matching and
constraint rewriting facilities which provide a finer and more
intuitive control for strategy definition on the other hand.


While our approach pursues the idea to integrate languages into a
system of cooperating solvers the approaches
\cite{Lopez:ALP1992,TOY-FD:1997,FHgSp:PADL2003,KobayashiMarinIda:2003,Lux:FLOPS2001}
come from the opposite point of view and extend the functional logic
program evaluation by constraint evaluation.

In contrast to the other approaches our framework allows the
integration of several constraint systems.  The user can
integrate desired domains and solvers which satisfy the interface
requirements as discussed in Sect.~\ref{sec:CooperatingSolvers}.
The opportunity to integrate different host languages and constraint
systems also distinguishes our approach from other existing systems of
cooperating solvers (for example \cite{Hong,Monfroy,Rueher}) that
usually have one fixed host language (a logic language).

Furthermore both, the above mentioned languages and the cooperative
systems, mainly have a fixed order of evaluation of constraints and
functional expressions resp.
In contrast, an integration according to our ideas using the \METASl
system allows the user to either define its own strategies or to
refine existing strategy patterns in a simple way.
The usefulness of different cooperation strategies has been proven for
usual solvers (\eg on arithmetic, see \cite{FHM03FLAIRS}).
As well, first results for the integration of a logic language into
\METASl confirm their usefulness for language solvers \cite{FHR:04}.

\Oz \cite{Oz:1995} supports (constraint) logic, functional and
object-oriented programming styles within one (as well fixed)
language.
The computation in \Oz is based on the concept of computation spaces
\cite{Schulte:2002} which consist of a constraint store containing
only basic constraints and propagators (for more complex constraints)
manipulating them.
Similar to our framework, computation spaces can be used to describe
solver cooperations and search strategies.  However, this relies on
all solvers sharing the same store format and hence is not satisfying
for the main goal of our approach, \ie the cooperation of black box
solvers independent of their implementation.

{\small\itshape%
  \textbf{Acknowledgement}. We want to thank our colleague Stephan Frank for
  many valuable comments and his assistance in preparing the manuscript. Our
  particular thanks goes to the reviewers who provided detailed and insightful
  comments that helped to improve the paper considerably.
}%




%% file: HofstedtPepper.bbl
\begin{thebibliography}{}

\bibitem[\protect\citeauthoryear{A{\"\i}t-Kaci and Nasr}{A{\"\i}t-Kaci and
  Nasr}{1989}]{AitKaciNasr:89}
{\sc A{\"\i}t-Kaci, H.} {\sc and} {\sc Nasr, R.} 1989.
\newblock {Integrating Logic and Functional Programming}.
\newblock {\em Lisp and Symbolic Computation\/}~{\em 2,\/}~1, 51--89.

\bibitem[\protect\citeauthoryear{Broy, Wirsing, and Pepper}{Broy
  et~al\mbox{.}}{1987}]{BroyWirsingPepper:87}
{\sc Broy, M.}, {\sc Wirsing, M.}, {\sc and} {\sc Pepper, P.} 1987.
\newblock {On the Algebraic Definition of Programming Languages}.
\newblock {\em ACM Transactions on Programming Languages and Systems\/}~{\em
  9,\/}~1 (January), 54--99.

\bibitem[\protect\citeauthoryear{Cheadle, Harvey, Sadler, Schimpf, Shen, and
  Wallace}{Cheadle et~al\mbox{.}}{2003}]{Eclipse:03}
{\sc Cheadle, A.~M.}, {\sc Harvey, W.}, {\sc Sadler, A.~J.}, {\sc Schimpf, J.},
  {\sc Shen, K.}, {\sc and} {\sc Wallace, M.~G.} 2003.
\newblock {\Eclipse. An Introduction}.
\newblock Tech. Rep. IC-PARC-03-1, Centre for Planning and Resource Control,
  Imperial College London.

\bibitem[\protect\citeauthoryear{Costa, Warren, and Yang}{Costa
  et~al\mbox{.}}{1991}]{CoWaYa:91}
{\sc Costa, V.~S.}, {\sc Warren, D.~H.}, {\sc and} {\sc Yang, R.} 1991.
\newblock {Andorra-I: A Parallel Prolog System that Transparently Exploits both
  And- and Or-Parallelism}.
\newblock In {\em Third ACM SIGPLAN Symposium on Principles and Practice of
  Parallel Programming -- PPOPP}. SIGPLAN Notices, vol. 26\,(7). ACM Press,
  83--93.

\bibitem[\protect\citeauthoryear{Didrich, Fett, Gerke, Grieskamp, and
  Pepper}{Didrich et~al\mbox{.}}{1994}]{Opal:94}
{\sc Didrich, K.}, {\sc Fett, A.}, {\sc Gerke, C.}, {\sc Grieskamp, W.}, {\sc
  and} {\sc Pepper, P.} 1994.
\newblock {OPAL: Design and Implementation of an Algebraic Programming
  Language}.
\newblock In {\em International Conference on Programming Languages and System
  Architectures}, {J.~Gutknecht}, Ed. LNCS, vol. 782. Springer, 228--244.

\bibitem[\protect\citeauthoryear{Fer\-n{\'a}ndez, Hortal{\'a}-Gonz{\'a}les, and
  S{\'a}enz-P{\'e}rez}{Fer\-n{\'a}ndez et~al\mbox{.}}{2003}]{FHgSp:PADL2003}
{\sc Fer\-n{\'a}ndez, A.}, {\sc Hortal{\'a}-Gonz{\'a}les, T.}, {\sc and} {\sc
  S{\'a}enz-P{\'e}rez, F.} 2003.
\newblock {Solving Combinatorial Problems with a Constraint Functional Logic
  Language}.
\newblock In {\em Practical Aspects of Declarative Languages, 5th International
  Symposium -- PADL}, {V.~Dahl} {and} {P.~Wadler}, Eds. LNCS, vol. 2562.
  Springer, 320--338.

\bibitem[\protect\citeauthoryear{Field and Harrison}{Field and
  Harrison}{1988}]{FieldHarrison:88}
{\sc Field, A.} {\sc and} {\sc Harrison, P.} 1988.
\newblock {\em {Functional Programming}}.
\newblock Addison-Wesley.

\bibitem[\protect\citeauthoryear{Frank, Hofstedt, and Mai}{Frank
  et~al\mbox{.}}{2003a}]{FHM03KI}
{\sc Frank, S.}, {\sc Hofstedt, P.}, {\sc and} {\sc Mai, P.~R.} 2003a.
\newblock A {F}lexible {M}eta-solver {F}ramework for {C}onstraint {S}olver
  {C}ollaboration.
\newblock In {\em Advances in Artificial Intelligence, 26th German Conference
  on Artificial Intelligence -- KI 2003}, {A.~G{\"u}nter}, {R.~Kruse}, {and}
  {B.~Neumann}, Eds. LNCS, vol. 2821. Springer, Kiel, Germany, 520--534.

\bibitem[\protect\citeauthoryear{Frank, Hofstedt, and Mai}{Frank
  et~al\mbox{.}}{2003b}]{FHM03FLAIRS}
{\sc Frank, S.}, {\sc Hofstedt, P.}, {\sc and} {\sc Mai, P.~R.} 2003b.
\newblock Meta-{S}: {A} {S}trategy-oriented {M}eta-{S}olver {F}ramework.
\newblock In {\em 16th {I}nternational {F}lorida {A}rtificial {I}ntelligence
  {R}esearch {S}ociety {C}onference -- {FLAIRS}}, {I.~Russell} {and} {S.~M.
  Haller}, Eds. The AAAI Press, St. Augustine, Florida, 177--181.

\bibitem[\protect\citeauthoryear{Frank, Hofstedt, and Reckmann}{Frank
  et~al\mbox{.}}{2004}]{FHR:04}
{\sc Frank, S.}, {\sc Hofstedt, P.}, {\sc and} {\sc Reckmann, D.} 2004.
\newblock {Strategies for the Efficient Solution of Hybrid Constraint Logic
  Programs}.
\newblock In {\em Third Workshop on Multiparadigm Constraint Programming
  Languages -- MultiCPL}, {S.~Mu{\~n}oz-Hern{\'a}ndez}, {J.~M.
  G{\'o}mez-Perez}, {and} {P.~Hofstedt}, Eds. Saint-Malo, France, 103--117.

\bibitem[\protect\citeauthoryear{Hanus}{Hanus}{1994}]{Hanus_FLP94}
{\sc Hanus, M.} 1994.
\newblock {The Integration of Functions into Logic Programming: From Theory to
  Practice}.
\newblock {\em Journal of Logic Programming\/}~{\em 19/20}, 583--628.

\bibitem[\protect\citeauthoryear{Hanus}{Hanus}{1995}]{Hanus:95}
{\sc Hanus, M.} 1995.
\newblock {On Extra Variables in (Equational) Logic Programming}.
\newblock In {\em Twelfth International Conference on Logic Programming}. MIT
  Press, 665--679.

\bibitem[\protect\citeauthoryear{Hanus, Antoy, Kuchen, L\'{o}pez-Fraguas, Lux,
  Moreno-Navarro, and Steiner}{Hanus et~al\mbox{.}}{2003}]{Curry:03}
{\sc Hanus, M.}, {\sc Antoy, S.}, {\sc Kuchen, H.}, {\sc L\'{o}pez-Fraguas,
  F.~J.}, {\sc Lux, W.}, {\sc Moreno-Navarro, J.~J.}, {\sc and} {\sc Steiner,
  F.} 2003.
\newblock {Curry: An Integrated Functional Logic Language}.
\newblock Report. Version 0.8 of April 15, 2003.

\bibitem[\protect\citeauthoryear{Hofstedt}{Hofstedt}{2000a}]{Hofstedt_CL2000}
{\sc Hofstedt, P.} 2000a.
\newblock {Better Communication for Tighter Cooperation}.
\newblock In {\em First International Conference on Computational Logic -- CL}.
  LNCS, vol. 1861. Springer, 342--358.

\bibitem[\protect\citeauthoryear{Hofstedt}{Hofstedt}{2000b}]{Hofstedt_CP2000}
{\sc Hofstedt, P.} 2000b.
\newblock {Cooperating Constraint Solvers}.
\newblock In {\em Sixth International Conference on Principles and Practice of
  Constraint Programming -- CP}. LNCS, vol. 1894. Springer, 520--524.

\bibitem[\protect\citeauthoryear{Hofstedt}{Hofstedt}{2001}]{Dissertation}
{\sc Hofstedt, P.} 2001.
\newblock {Cooperation and Coordination of Constraint Solvers}.
\newblock Ph.D. thesis, Dresden University of Technology.

\bibitem[\protect\citeauthoryear{Hong}{Hong}{1994}]{Hong}
{\sc Hong, H.} 1994.
\newblock {Confluency of Cooperative Constraint Solvers}.
\newblock Tech. Rep. 94--08, Research Institute for Symbolic Computation, Linz,
  Austria.

\bibitem[\protect\citeauthoryear{Hortal\'{a}-Gonz\'{a}lez, L\'{o}pez-Fraguas,
  S\'{a}nchez-Hern\'{a}ndez, and
  Ull\'{a}n-Hern\'{a}ndez}{Hortal\'{a}-Gonz\'{a}lez
  et~al\mbox{.}}{1997}]{TOY-FD:1997}
{\sc Hortal\'{a}-Gonz\'{a}lez, T.}, {\sc L\'{o}pez-Fraguas, F.}, {\sc
  S\'{a}nchez-Hern\'{a}ndez, J.}, {\sc and} {\sc Ull\'{a}n-Hern\'{a}ndez, E.}
  1997.
\newblock {Declarative Programming with real Constraints}.
\newblock Tech. Rep. SIP 5997, Universidad Complutense de Madrid.

\bibitem[\protect\citeauthoryear{Hudak, Peterson, and Fasel}{Hudak
  et~al\mbox{.}}{2000}]{Haskell:99}
{\sc Hudak, P.}, {\sc Peterson, J.}, {\sc and} {\sc Fasel, J.~H.} 2000.
\newblock {A Gentle Introduction to {Haskell 98}}.
\newblock Tech. rep., Yale University, Department of Computer Science.
\newblock \url{http://www.haskell.org/tutorial/}, visited 2005-11-18.

\bibitem[\protect\citeauthoryear{Jaffar and Lassez}{Jaffar and
  Lassez}{1987}]{JaffarLassezCLP}
{\sc Jaffar, J.} {\sc and} {\sc Lassez, J.-L.} 1987.
\newblock {Constraint Logic Programming}.
\newblock In {\em 14th ACM Symposium on Principles of Programming Languages --
  POPL}. ACM Press, 111--119.

\bibitem[\protect\citeauthoryear{Jaffar, Maher, Marriott, and Stuckey}{Jaffar
  et~al\mbox{.}}{1998}]{JaffarMaherMarriottStuckey}
{\sc Jaffar, J.}, {\sc Maher, M.}, {\sc Marriott, K.}, {\sc and} {\sc Stuckey,
  P.} 1998.
\newblock {The Semantics of Constraint Logic Programs}.
\newblock {\em Journal of Logic Programming\/}~{\em 37,\/}~1-3, 1--46.

\bibitem[\protect\citeauthoryear{Kanellakis, Kuper, and Revesz}{Kanellakis
  et~al\mbox{.}}{1995}]{ConstraintDB:1995}
{\sc Kanellakis, P.}, {\sc Kuper, G.}, {\sc and} {\sc Revesz, P.} 1995.
\newblock {Constraint Query Languages}.
\newblock {\em Journal of Computer and System Sciences\/}~{\em 51,\/}~1,
  26--52.

\bibitem[\protect\citeauthoryear{Kobayashi, Marin, and Ida}{Kobayashi
  et~al\mbox{.}}{2003}]{KobayashiMarinIda:2003}
{\sc Kobayashi, N.}, {\sc Marin, M.}, {\sc and} {\sc Ida, T.} 2003.
\newblock {Collaborative Constraint Functional Logic Programming System in an
  Open Environment}.
\newblock {\em IEICE Transactions on Information and Systems\/}~{\em
  E86-D,\/}~1, 63--70.

\bibitem[\protect\citeauthoryear{Loogen}{Loogen}{1995}]{Loogen:95}
{\sc Loogen, R.} 1995.
\newblock {\em Integration funktionaler und logischer Programmiersprachen}.
\newblock Oldenbourg.

\bibitem[\protect\citeauthoryear{L{\'o}pez-Fraguas}{L{\'o}pez-Fraguas}{1992}]{%
Lopez:ALP1992}
{\sc L{\'o}pez-Fraguas, F.-J.} 1992.
\newblock {A General Scheme for Constraint Functional Logic Programming}.
\newblock In {\em Algebraic and Logic Programming -- ALP}, {H.~Kirchner} {and}
  {G.~Levi}, Eds. LNCS, vol. 632. Springer, 213--227.

\bibitem[\protect\citeauthoryear{Lux}{Lux}{2001}]{Lux:FLOPS2001}
{\sc Lux, W.} 2001.
\newblock {Adding Linear Constraints over Real Numbers to Curry}.
\newblock In {\em Functional and Logic Programming, 5th International Symposium
  -- FLOPS}, {H.~Kuchen} {and} {K.~Ueda}, Eds. LNCS, vol. 2024. Springer,
  185--200.

\bibitem[\protect\citeauthoryear{Manna}{Manna}{1974}]{Manna:74}
{\sc Manna, Z.} 1974.
\newblock {\em {Mathematical Theory of Computation}}.
\newblock McGraw-Hill.

\bibitem[\protect\citeauthoryear{Middeldorp and Hamoen}{Middeldorp and
  Hamoen}{1994}]{MidHam:94}
{\sc Middeldorp, A.} {\sc and} {\sc Hamoen, E.} 1994.
\newblock {Completeness Results for Basic Narrowing}.
\newblock {\em Applicable Algebra in Engineering, Communication and
  Computing\/}~{\em 5}, 213--253.

\bibitem[\protect\citeauthoryear{Monfroy}{Monfroy}{1996}]{Monfroy}
{\sc Monfroy, E.} 1996.
\newblock {Solver Collaboration for Constraint Logic Programming}.
\newblock Ph.D. thesis, Centre de Recherche en Informatique de Nancy. INRIA.

\bibitem[\protect\citeauthoryear{Moreno-Navarro and
  Rodr\'{\i}guez-Artalejo}{Moreno-Navarro and
  Rodr\'{\i}guez-Artalejo}{1992}]{Babel:1992}
{\sc Moreno-Navarro, J.} {\sc and} {\sc Rodr\'{\i}guez-Artalejo, M.} 1992.
\newblock {Logic Programming with Functions and Predicates: The Language
  BABEL}.
\newblock {\em Journal for Logic Programming\/}~{\em 12,\/}~3{\&}4, 191--223.

\bibitem[\protect\citeauthoryear{M\"uller, M\"uller, and Van~Roy}{M\"uller
  et~al\mbox{.}}{1995}]{Oz:1995}
{\sc M\"uller, M.}, {\sc M\"uller, T.}, {\sc and} {\sc Van~Roy, P.} 1995.
\newblock {Multiparadigm Programming in {Oz}}.
\newblock In {\em Workshop on Visions for the Future of Logic Programming},
  {D.~Smith}, {O.~Ridoux}, {and} {P.~V. Roy}, Eds. Portland, Oregon.

\bibitem[\protect\citeauthoryear{Nilsson and Ma{\l}uszy\'{n}ski}{Nilsson and
  Ma{\l}uszy\'{n}ski}{1995}]{Prolog}
{\sc Nilsson, U.} {\sc and} {\sc Ma{\l}uszy\'{n}ski, J.} 1995.
\newblock {\em {Logic, Programming and Prolog}}.
\newblock John Wiley \& Sons Ltd.

\bibitem[\protect\citeauthoryear{Palamidessi}{Palamidessi}{1990}]{Palamidessi}
{\sc Palamidessi, C.} 1990.
\newblock {Algebraic Properties of Idempotent Substitutions}.
\newblock In {\em Automata, Languages and Programming -- ICALP}, {M.~Paterson},
  Ed. LNCS, vol. 443. Springer, 386--399.

\bibitem[\protect\citeauthoryear{Reddy}{Reddy}{1985}]{Reddy:85}
{\sc Reddy, U.} 1985.
\newblock {Narrowing as the Operational Semantics of Functional Languages}.
\newblock In {\em IEEE Symposium on Logic Programming}. 138--151.

\bibitem[\protect\citeauthoryear{Rueher}{Rueher}{1995}]{Rueher}
{\sc Rueher, M.} 1995.
\newblock {An Architecture for Cooperating Constraint Solvers on Reals}.
\newblock In {\em Constraint Programming: Basics and Trends}, {A.~Podelski},
  Ed. LNCS, vol. 910. Springer, 231--250.

\bibitem[\protect\citeauthoryear{Schulte}{Schulte}{2002}]{Schulte:2002}
{\sc Schulte, C.} 2002.
\newblock {\em Programming Constraint Services}. LNCS, vol. 2302.
\newblock Springer.

\bibitem[\protect\citeauthoryear{Suzuki, Middeldorp, and Ida}{Suzuki
  et~al\mbox{.}}{1995}]{SMI:95}
{\sc Suzuki, T.}, {\sc Middeldorp, A.}, {\sc and} {\sc Ida, T.} 1995.
\newblock {Level-Confluence of Conditional Rewrite Systems with Extra Variables
  in Right-Hand Sides}.
\newblock In {\em 6th International Conference on Rewriting Techniques and
  Applications}. LNCS, vol. 914. Springer, 179--193.

\bibitem[\protect\citeauthoryear{Warren}{Warren}{1988}]{Warren:88}
{\sc Warren, D.~H.} 1988.
\newblock {The Andorra Principle}.
\newblock Presented at the Gigalips Workshop, Swedish Institute of Computer
  Science (SICS), Stockholm, Sweden.

\bibitem[\protect\citeauthoryear{Westfold and Smith}{Westfold and
  Smith}{2001}]{WesSmi:01}
{\sc Westfold, S.~J.} {\sc and} {\sc Smith, D.~R.} 2001.
\newblock {Synthesis of Efficient Constraint Satisfaction Programs}.
\newblock In {\em Knowledge Engineering Reviews. Special Issue on AI and OR}.
  Kestrel Institute Technical Report KES.U.01.7.

\bibitem[\protect\citeauthoryear{Winskel}{Winskel}{1993}]{Winskel:93}
{\sc Winskel, G.} 1993.
\newblock {\em {Formal Semantics of Programming Languages}}.
\newblock MIT Press.

\end{thebibliography}
